\documentclass[prx,twocolumn,superscriptaddress]{revtex4-1}
\usepackage{amsmath}
\usepackage{amssymb}
\usepackage{amsthm}
\usepackage{amsfonts}
\usepackage{listings}
\usepackage{diagbox}
\usepackage{enumerate}
\usepackage{latexsym}
\usepackage{color}
\usepackage{xcolor}
\usepackage{bm}
\usepackage{hyperref}
\usepackage{graphicx}
\usepackage[FIGTOPCAP]{subfigure}
\hypersetup{
	pdfnewwindow=true, colorlinks=true,
	linkcolor=blue, anchorcolor=blue,
	citecolor=blue, filecolor=blue,
	menucolor=blue, urlcolor=blue}

\newcommand{\beginsupplement}{%
	\setcounter{table}{0}
	\renewcommand{\thetable}{S\arabic{table}}%
	\setcounter{figure}{0}
	\renewcommand{\thefigure}{S\arabic{figure}}%
}

\newcommand{\ket}[1]{\left|#1\right\rangle}

\def\ie{{\it i.e.},\ }
\def\eg{{\it e.g.}\ }
\def\ea{{\it et al.}}

\input{epsf}

\begin{document}

\tolerance 10000

\draft

\title{Unprotected quadratic band crossing points and quantum anomalous Hall effect in FeB$_2$ monolayer}

\author{Dongyu Wu}
\thanks{These authors contributed equally to this work.}
\affiliation{Beijing National Laboratory for Condensed Matter Physics,
	and Institute of Physics, Chinese Academy of Sciences, Beijing 100190, China}
\affiliation{University of Chinese Academy of Sciences, Beijing 100049, China}

\author{Yunpeng Huang}
\thanks{These authors contributed equally to this work.}
\affiliation{Beijing National Laboratory for Condensed Matter Physics,
	and Institute of Physics, Chinese Academy of Sciences, Beijing 100190, China}
\altaffiliation{These authors contributed equally to this work.}

\author{Song Sun}
\affiliation{Beijing National Laboratory for Condensed Matter Physics,
	and Institute of Physics, Chinese Academy of Sciences, Beijing 100190, China}
\affiliation{University of Chinese Academy of Sciences, Beijing 100049, China}

\author{Jiacheng Gao}
\affiliation{Beijing National Laboratory for Condensed Matter Physics,
	and Institute of Physics, Chinese Academy of Sciences, Beijing 100190, China}
\affiliation{University of Chinese Academy of Sciences, Beijing 100049, China}

\author{Zhaopeng Guo}
\email{zpguo@iphy.ac.cn}
\affiliation{Beijing National Laboratory for Condensed Matter Physics,
	and Institute of Physics, Chinese Academy of Sciences, Beijing 100190, China}

\author{Hongming Weng}
\affiliation{Beijing National Laboratory for Condensed Matter Physics,
	and Institute of Physics, Chinese Academy of Sciences, Beijing 100190, China}
\affiliation{University of Chinese Academy of Sciences, Beijing 100049, China}

\author{Zhong Fang}
\affiliation{Beijing National Laboratory for Condensed Matter Physics,
	and Institute of Physics, Chinese Academy of Sciences, Beijing 100190, China}
\affiliation{University of Chinese Academy of Sciences, Beijing 100049, China}

\author{Kun Jiang}
\affiliation{Beijing National Laboratory for Condensed Matter Physics,
	and Institute of Physics, Chinese Academy of Sciences, Beijing 100190, China}

\author{Zhijun Wang}
\email{wzj@iphy.ac.cn}
\affiliation{Beijing National Laboratory for Condensed Matter Physics,
	and Institute of Physics, Chinese Academy of Sciences, Beijing 100190, China}
\affiliation{University of Chinese Academy of Sciences, Beijing 100049, China}

\date{\today}

\begin{abstract}
Quadratic band crossing points (QBCPs) and quantum anomalous Hall effect (QAHE) have attracted the attention of both theoretical and experimental researchers in recent years. Based on first-principle calculations, we find that the FeB$_2$ monolayer is a nonmagnetic semimetal with QBCPs at $K$. Through symmetry analysis and $\mathbf{k}\cdot\mathbf{p}$ invariant theory, we find that the QBCP is not protected by rotation symmetry and consists of two Dirac points with same chirality (Berry phase of $2\pi$).
Once introducing Coulomb interactions, we find that there is a spontaneous-time-reversal-breaking instability of the spinful QBCPs, which gives rise to a $C=2$ QAH insulator with orbital moment ordering.
\end{abstract}

\maketitle

\section{Introduction}
In a two-dimensional (2D) system, the finite density of states associated with the parabolic dispersion could lead to instability for arbitrarily weak interactions~\cite{quadratic2009,quadratic2020,EPJB2012,quadratic2010,liang_interaction-driven_2017}.
For a quadratic band crossing point (QBCP) being stable without fine-tuning, the system must be time-reversal invariant and the QBCP must have $C_4$ or $C_6$ rotational symmetry~\cite{quadratic2009}. An interaction would lead to the possibility of spontaneous breaking of rotational symmetry (nematic phase) or time-reversal invariance.
However, the QBCPs at the threefold-invariant momentum on the honeycomb lattice and relatives are unprotected.
The introduction of interactions leads to qualitatively different low-energy behavior without breaking the underlying symmetries~\cite{quadratic2020}.
Although there are many theoretical studies of spinless QBCPs on many tailored 2D systems, such as single-layer graphene~\cite{quadratic2020,EPJB2012} and Bernal-stacked bilayer graphene~\cite{quadratic2010}, the unprotected QBCPs have not been reported in any spinful system and their possible instabilities have not been discussed yet.

Besides, topological states, including quantum anomalous Hall (QAH) state, have attracted considerable research interest recently~\cite{QSH-2006,Bi2Se3-2009,TaSe3_2018,Weyl-2009,TaAs-2015-Weng,TaAs-2015-Xu,ZrCo2Sn2016,TKNN}.
In spite of plenty of material proposals for QAH state~\cite{Yu2010,Qiao2010,PhysRevLett.110.116802,Xue2018,PhysRevLett.110.196801,EuB6_2020}, the observation of the QAH effect is still full of challenges and has been merely realized in a few systems such as, Cr-doped and V-doped (Bi,Sb)$_2$Te$_3$ thin films~\cite{Chang2013,Chang2015}, magnetic topological insulator MnBi$_2$Te$_4$~\cite{QAH-MnBi2Te4} and twisted bilayer graphene (TBG)~\cite{QAH-TBG}. Previous theoretical studies also show that the QAH effect can be realized in graphene by introducing both exchange field and Rashba spin-orbit coupling (SOC) due to its unique linear Dirac band dispersions~\cite{Qiao2010}. 

In recent years, $M$B$_2$ ($M=$ transition metal) monolayers have been predicted to be 2D Dirac cone materials theoretically in the absence of SOC, such as TiB$_2$~\cite{TiB2-2014}, FeB$_2$~\cite{FeB2-2016} and HfB$_2$~\cite{HfB2-2019} monolayers.
The FeB$_2$ bulk crystal has been grown~\cite{FeB2-1970}. The stability of the FeB$_2$ monolayer is predicted theoretically by structure searching, phonon spectra, and molecular dynamics~\cite{FeB2-2016,FeB2-2018,FeB2-2019}.
Unlike graphene, the Dirac bands of FeB$_2$ originate from $d$ states of the transition metal Fe, which has a substantial Rashba SOC effect and is very likely coupled to a magnetic field. A Chern insulator can be achieved once it is grown on an insulating magnetic substrate. 
In this article, we find that the FeB$_2$ monolayer is a nonmagnetic semimetal with QBCPs based on first-principle calculations.
Without including SOC, there is a linearly dispersive Dirac node at $K$ (resp. $K'$) with a Berry phase $\pi$ (resp. $-\pi$), protected by the combined symmetry of time reversal and twofold rotation (\ie $TC_{2z}$). Once including SOC, the Dirac node becomes a QBCP and its Berry phase becomes $2\pi~(-2\pi)$ at $K~(K')$.
These characters can be captured by the $\mathbf{k}\cdot\mathbf{p}$ effective Hamiltonians.
With an insulating magnetic substrate, the FeB$_2$ monolayer is turned to be a Chern insulator with two chiral edge states, which is stimulated by the fixed-moment calculations. More interestingly, once we introduce the Coulomb interactions, an instability towards a $C=2$ QAH state with orbital moment ordering is found.

\section{Calculation method}
We performed the first-principles calculations within the framework of the density functional theory (DFT) using the projector augmented wave (PAW) method~\cite{PAW1994,PAW1999}, which is implemented in Vienna \emph{ab initio} simulation package (VASP)~\cite{VASP1,VASP2}.
The Perdew-Burke-Ernzerhof (PBE) generalized gradient approximation exchange-correlation functional~\cite{PBE} was implemented in calculations.
The cut-off energy for plane wave expansion was 945 eV, and 12 $\times$ 12 $\times$ 1 $k$-point sampling grids were used in the self-consistent process.
A vacuum layer of 20~\AA~was chosen to avoid interaction between neighboring layers.
SOC was taken into account within the second variational method self-consistently.
The irreducible representations (irreps) were obtained by the program IRVSP~\cite{Irvsp}.
The maximally localized Wannier functions (MLWFs) were constructed by Fe-$3d$, B-$2s$ and B-$2p$ orbitals using Wannier90 package~\cite{wannier90}. The edge states were calculated using surface Green's function of the semi-infinite system based on the iterative scheme~\cite{Green1984,Green1985,wanniertools}.

\begin{figure}[!t]
	\includegraphics[width=3.2in]{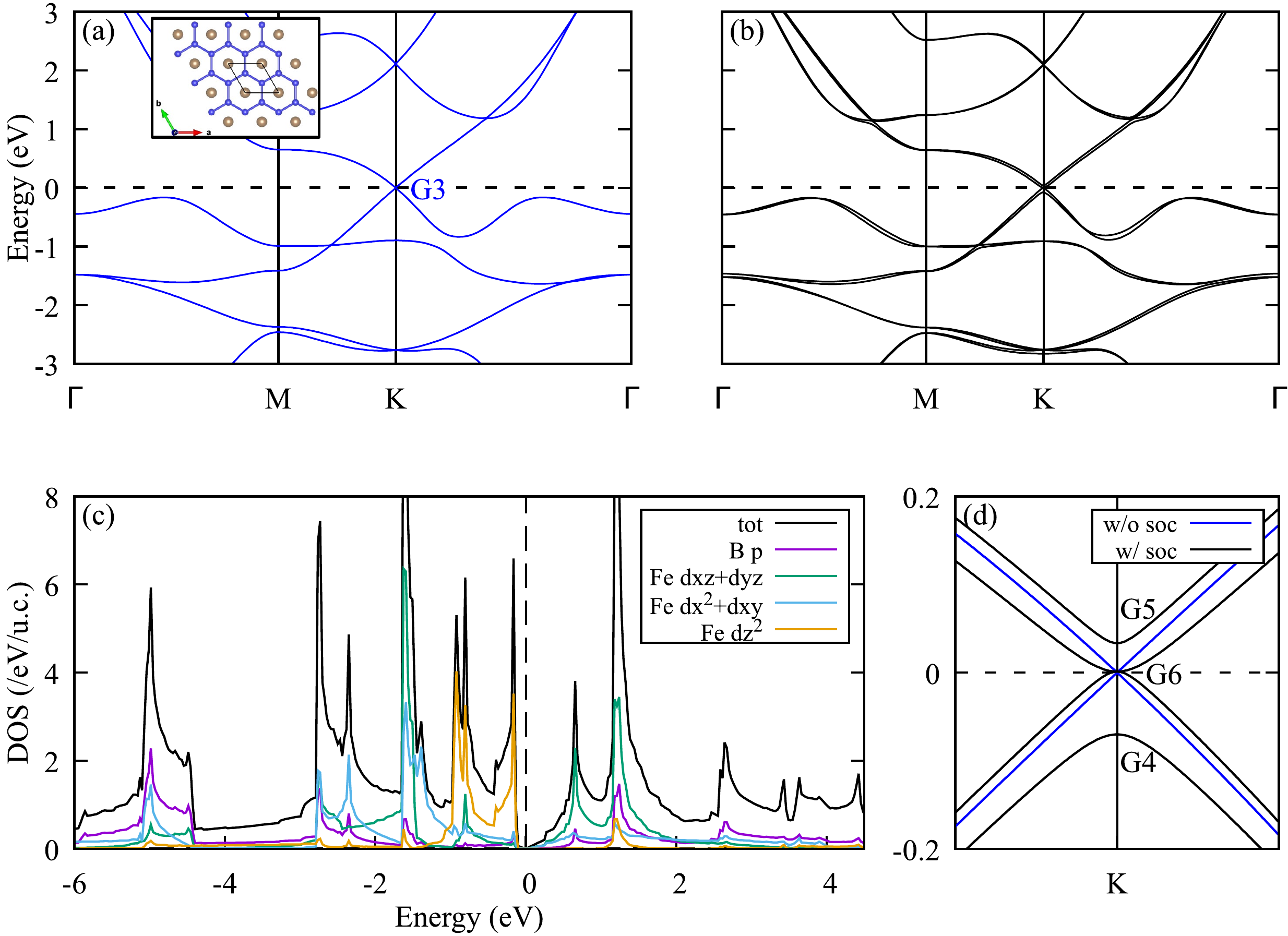}
	\caption{(color online).
		The calculated electronic structures of FeB$_2$ monolayer without (a) and with (b) SOC. 
		The inset in (a) shows the crystal structure of the FeB$_2$ monolayer, where Fe and B atoms are marked by brown and blue balls.
		The irrep of two-fold bands at $K$ is denoted by $G_{3}$.
		(c) The total DOS and projected DOS of B-$p$ and Fe-$d$ orbitals, respectively. 
		(d) The zoom-in plot of band structures near $K$ point. The irreps of four low-energy SOC bands at $K$ are denoted by $G_{4}$, $G_{5}$, and $G_{6}$.
	}\label{fig:dft}
\end{figure}

\section{DFT results}
The crystal structure of FeB$_2$ monolayer belongs to space group $P6mm$ (No. 183), as shown in Fig.~\ref{fig:dft}(a).
One unit cell (a = 3.171~\AA) contains two B atoms and one Fe atom, which are located at $2b$ and $1a$ Wyckoff positions, respectively.
B atoms are arranged in a honeycomb lattice, and Fe atoms are located in the middle of the hexagons.
The distance of Fe atoms and B atoms plane is 0.628~\AA.
The PBE band structure of FeB$_2$ monolayer is shown in Fig.~\ref{fig:dft}(a), there is a linearly dispersive Dirac point at $K$ near the Fermi level ($E_F$).
Therefore, FeB$_2$ monolayer was predicted to be a 2D Dirac semimetal~\cite{FeB2-2016,FeB2-2020}.
The little point group at $K$ is C$_{3v}$, and the twofold Dirac bands belong to $G_{3}$ irreducible representation (irrep), which is consistent with Ref.~\cite{FeB2-2020}.
The total and projected density of states (DOS) are plotted in Fig.~\ref{fig:dft}(c). They show that the hybridization between Fe and B is strong, while the electronic states near $E_F$ are mainly contributed by $d_{z^2}$ and $d_{xz}+d_{yz}$ electrons of Fe atoms (the orbital-resolved band structures are given in Appendix~\ref{sup:A}).
Once including SOC, as shown in Figs.~\ref{fig:dft}(b,d), the two Dirac bands split into two non-degenerate bands ($G_4$ and $G_5$) and one doubly-degenerate band ($G_6$) at $K$, exhibiting quadratic band dispersion.

\section{Low-energy effective models}
Based on the theory of invariants, we derive the low-energy effective Hamiltonian $H_K(\vec{k})$ (\ie$\vec{k}$ is the offset momentum from $K$). Under the basis of $G_3$ irrep,~\eg $\{\ket{d_{xz}+id_{yz}},\ket{d_{xz}-id_{yz}}\}$, it reads,
\begin{equation}\label{eq:kp-nsoc}
	\begin{split}
		H_{K}(\vec{k})&=
		\left(
		\begin{array}{cc}   %
			M_{1}(\vec{k}) &  Ak_{+}  \\ 
			Ak_{-} & M_{1}(\vec{k})  \\
		\end{array}
		\right).
	\end{split}
\end{equation}
After considering the spin degree of freedom, the four-band Hamiltonian becomes (in the basis of $\{\ket{\uparrow} , \ket{\downarrow}\} \otimes \{\ket{d_{xz}+id_{yz}}, \ket{d_{xz}-id_{yz}}\}$),
\begin{equation}\label{eq:kp-soc}
	\begin{split}
		H^{so}_{K}(\vec{k})&=\sigma_0\otimes H_{K}(\vec{k}) \\
		&+	\left(
		\begin{array}{cccc}   %
			M_{2}(\vec{k}) &          & &\dagger      \\  %
			0  &  -M_{2}(\vec{k})   & &\\  %
			iBk_{+}   &   iC(\vec{k})   &  -M_{2}(\vec{k})  &  \\  %
			iM_3(\vec{k})  &    iBk_{+}    & 0 &  M_{2}(\vec{k})  \\  %
		\end{array}
		\right),
	\end{split}
\end{equation}
where $k_{\pm}=k_{x}\pm ik_{y}$, $C(\vec{k})=C_{1}k_{-}+C_{2}k_{+}^{2}$, $M_{\alpha=1,2,3}(\vec{k})=E_{\alpha}+F_{\alpha}k_{\bot}^2$ with $k_{\bot}^{2}=k_{x}^{2}+k_{y}^{2}$, and $\sigma_0$ is the identify matrix in spin space.
Since the second-order $k$ terms are crucial for the quadratic band dispersion in the following two-band model (which were omitted in Ref.~\cite{FeB2-2020}), we have derived them in the four-band $\mathbf{k}\cdot\mathbf{p}$ model as well.

\begin{figure}[!t]
	\includegraphics[width=3.2in]{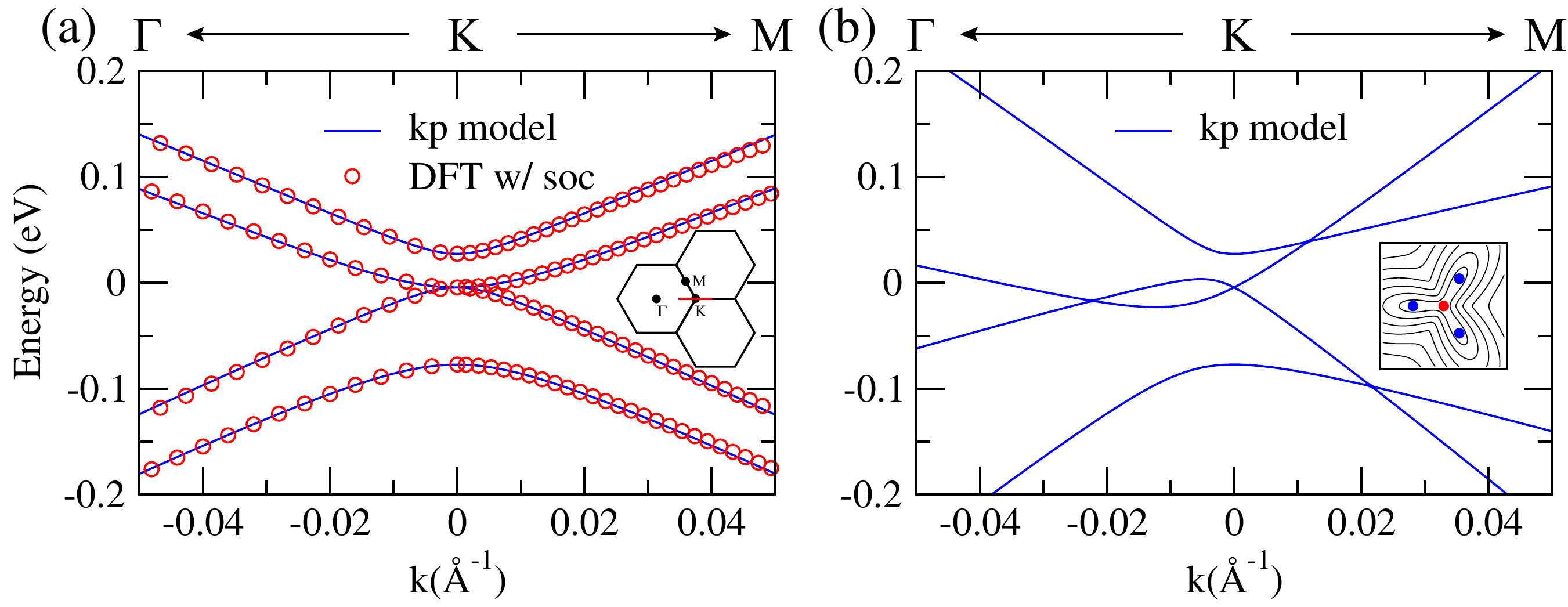}
	\caption{(color online).
		(a) The band dispersions (blue lines) of the four-band effective Hamiltonian in Eq.~(\ref{eq:kp-soc}) with the parameters in Table~\ref{table:kp} agree well with the bands (red circles) from the DFT calculation. The k-path is indicated by the red line in the inset of the Brillouin zone. 
		(b) The band dispersions of the effective model with $C_1=-3.03 $ eV\AA~and $C_2=-3.03$~eV\AA$^2$ are plotted for comparison.
		The inset shows iso-energy-gap lines in the vicinity of $K$ point. The Dirac points with Berry phase $\pi$ and $-\pi$ are colored in blue and red, respectively.
	}\label{fig:kp}
\end{figure}

\begin{table}[!b]
	\setlength{\tabcolsep}{1.0mm}
	\setlength{\extrarowheight}{2pt}
	\caption{The parameters in the effective $\mathbf{k}\cdot\mathbf{p}$ Hamiltonian.}
	\begin{tabular}{c c c c c c}%
		\hline
		\hline
		0th order  & eV & 1st order & eV$\cdot${\AA} & 2nd order & eV$\cdot${\AA}$^{2}$   \\
		\hline 
		$E_{1}$ & -0.0147 & $A$ & 2.6021 & $F_{1}$ & -1.7201\\
		$E_{2}$ & -0.0103 & $B$ & 0.0141 & $F_{2}$ & 0.2346\\
		$E_{3}$ & 0.0522 & $C_{1}$ & -0.0080 & $F_{3}$ & 0.5754\\
		&  &   &  & $C_{2}$ & -0.1010\\
		\hline
		\hline
	\end{tabular}
	\label{table:kp}
\end{table}

By fitting the DFT band structure in the vicinity of $K$, the parameters are obtained in Table~\ref{table:kp} and the results are shown in Fig.~\ref{fig:kp}(a). The $\mathbf{k}\cdot\mathbf{p}$ model reproduces the QBCP at $K$, and the Berry phase for the QBCP is $2\pi$. 
The $2nd$ and $3rd$ bases form $G_6$ irrep of $C_{3v}$ double group, \ie$\{\ket{d_{xz}-id_{yz},\uparrow},\ket{d_{xz}+id_{yz},\downarrow}\} $. 
To evaluate the positions of Dirac points, a simple model under $G_6$ irrep can be obtained as below,
\begin{equation}\label{eq:kp-g6}
	\begin{split}
		H'_{K}(\vec{k})&=
		\left(
		\begin{array}{cc}   %
			M_{1}(\vec{k})-M_{2}(\vec{k})   &  -iC_{1}k_{+}-iC'_{2}k_{-}^{2} \\  %
			iC_{1}k_{-}+iC'_{2}k_{+}^{2}   &   M_{1}(\vec{k})-M_{2}(\vec{k})  \\  %
		\end{array}
		\right),
	\end{split}
\end{equation}
where $C'_{2}$ is a modified parameter after downfolding.
Its two eigenvalues are solved as $E_{\pm}=M_{1}(\vec{k})-M_{2}(\vec{k})  \pm \sqrt{\Delta(\vec{k})}$ with
\begin{equation}
	\Delta(\vec{k})={C'_{2}}^2k_{\bot}^4+2C_{1}C'_{2}k_x(k_x^2-3k_y^2)+{C_{1}}^2k_{\bot}^2.
\end{equation}
The gapless points satisfy the condition of $\Delta(\vec{k})=0$. 
Assuming  $k_y=0$, the equation is simplified to ${{C'_{2}}^2k_x^4+2C_{1}C'_{2}k_x^3+{C_{1}}^2k_x^2}=0$, giving rise to two Dirac points located at $k_x=0$, $-C_{1}/C'_{2}$. The detailed calculations show that the two Dirac points have opposite $\pi$ Berry phase. 
The distance between them is $d_0=|C_{1}/C'_{2}|$ in momentum space.
Considering $C_{3z}$ symmetry, there must be two additional Dirac points around K, as shown in the inset of Fig.~\ref{fig:kp}(b). No other gapless point is found (see the proof in Appendix~\ref{sup:B}).

\begin{figure}[t]
	\includegraphics[width=3.2in]{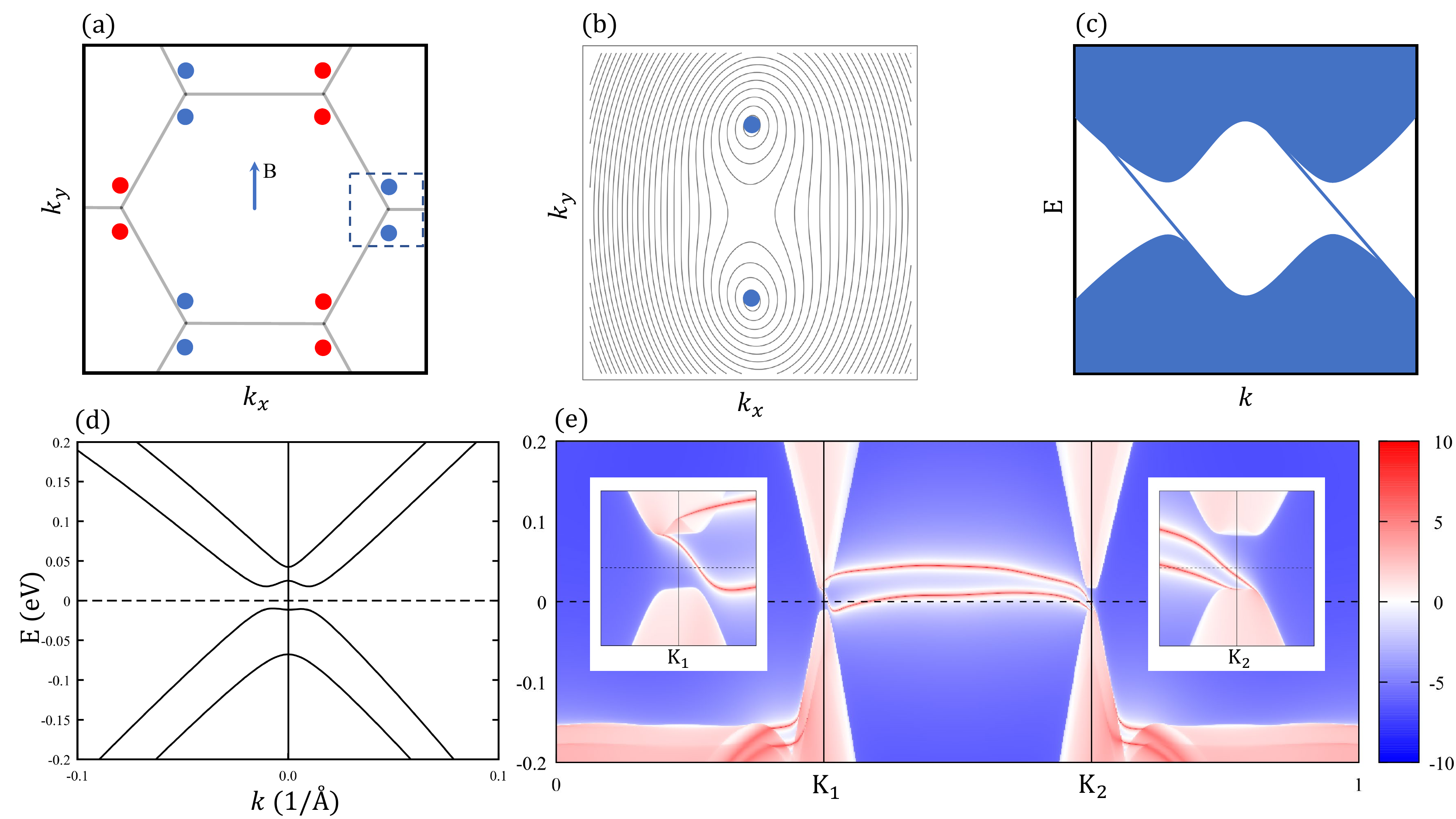}
	\caption{(color online).
		The electronic structure of FeB$_2$ monolayer in a magnetic field. 
		(a) The distribution of Dirac points with the in-plane magnetic field (which keeps $TC_{2z}$). The arrow marks the direction of the magnetic field. Each QBCP splits into two Dirac points with identical $\pi$ Berry phase.
		(b) The iso-energy-gap contours are plotted in the dashed rectangle area in (a).
		(c) The schematic diagram of two chiral edge states of FeB$_2$ monolayer with $z$-directed magnetic field.
		(d) The computed band structure of FeB$_2$ monolayer with a fixed magnetic moment 0.01~$\mu_B$ on each Fe atom.
		(e) The B-terminated zigzag-edge state of fixed magnetic moment FeB$_2$ monolayer. The inserts show the zoom-in plots around K$_{1,2}$. The K$_1$ and K$_2$ are the projections of $K$ and $K'$ on the edge.
	}\label{fig:mag}
\end{figure}

As the Dirac points and the quantized Berry phase of $\pi$ are protected by the antiunitary symmetry $TC_{2z}$, the above discussion should be valid for the four-band model $H_K^{so}(\vec{k})$ as well. 
In the band dispersions of Fig.~\ref{fig:kp}(a), we numerically get $d_0\sim5.2\times10^{-5}$~\AA$^{-1}$. The ration $d_0/d_{\Gamma K}$ is 0.004\% ($d_{\Gamma K}=1.321$~\AA$^{-1}$), which is too small to identify in FeB$_2$ monolayer. Therefore, it's rational to consider the $K$ point is a double Dirac point with quadratic band dispersions in FeB$_2$ monolayer, corresponding to a $2\pi$ Berry phase. As the quadratic band dispersion is not protected by rotational symmetry, it was previously considered as linear dispersion improperly~\cite{FeB2-2020}. Note that it's similar to the case in the magic-angle TBG, where the velocity of $K$ becomes zero~\cite{TBG2011,PhysRevLett.123.036401}.
For comparison, we plot the band dispersions of the four-band model with different $C_1$ and $C_2$ parameters in Fig.~\ref{fig:kp}(b), from which $d_0$ is read to be $0.223$~\AA$^{-1}$ ($\sim0.17d_{\Gamma K}$). The iso-energy-gap contours are shown in the vicinity of $K$ points in its inset. A Dirac point ($-\pi$) at $K$ and three other Dirac points ($\pi$) are clearly shown.

With an external magnetic field, a Chern insulator can be achieved in FeB$_2$ monolayer ({\it e.g.}, grown on an insulating magnetic substrate). As shown in Figs.~\ref{fig:mag}(a,b), with an in-plane external magnetic field (keeping $TC_{2z}$), the double Dirac point at $K$ splits into two Dirac points with the same chirality. The positions of Dirac nodes with different strength and directions of the in-plane magnetic field are shown in Fig.\ref{fig:mag_in_plane}. When the magnetism is out-of-plane, the FeB$_2$ becomes a Chern insulator with two chiral edge states in Fig.~\ref{fig:mag}(c). The Zeeman's coupling Hamiltonian is given in Appendix~\ref{sup:C}.
To simulate the spin-polarized state of FeB$_2$ induced by the out-of-plane magnetism of substrates, we have performed the DFT calculations with a fixed moment (\eg 0.01~$\mu_B$ on each Fe atom) in $z$ direction. Its spin-polarized band structure is obtained in Fig.~\ref{fig:mag}(d). The FeB$_2$ monolayer becomes a Chern insulator, which is compatible with the result of graphene with both Rashba SOC and an exchange field~\cite{Qiao2010}. Then, we construct the maximally localized Wannier functions (MLWFs) and calculate the edge spectra, as shown in Fig.~\ref{fig:mag}(e). Two chiral edge states connecting the conduction continuum and valence continuum indicate a Chern number of 2. 

\begin{figure}[t]
	\begin{center}
		\includegraphics[width=3.2in]{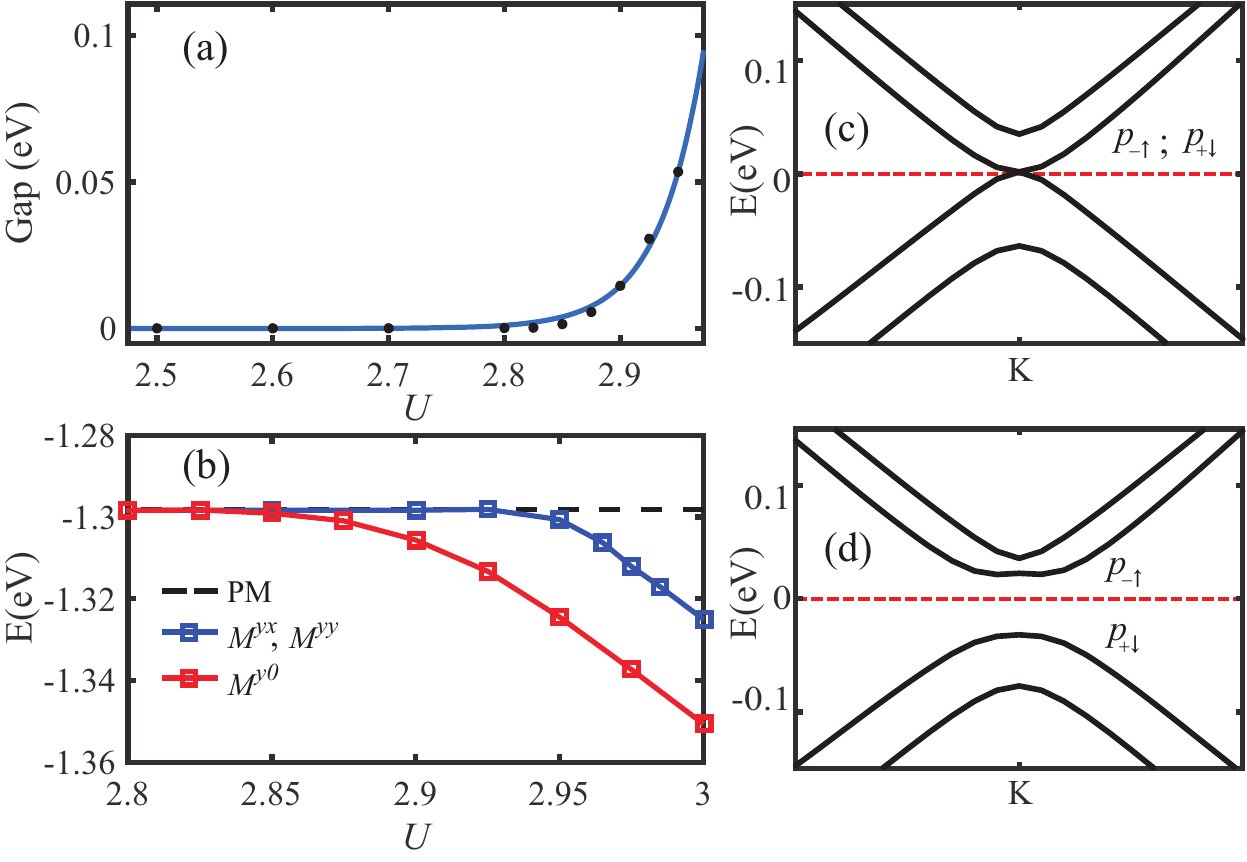}
		\caption{(a) The energy gap behavior of the $M^{y0}$ phase when varying the Coulomb interaction strength $U$, the solid blue line is fitting to the data by exponential function $\Delta=a*\exp\left(-1/\left(b*U\right)\right)$ with $a=3.812\times10^{30}$ and $b=4.622\times10^{-3}$. (b) The energy of the obtained states in the self-consistency calculation as a function of the interaction strength $U$. (c-d) The quasiparticle energy bands when $U=0$ and $U=2.95~eV$, the orbital characters of QBCP are also shown. The basis $P_{\pm\alpha}=\frac{1}{\sqrt{2}}\left(d_{xz,\alpha}\pm i d_{yz,\alpha}\right)$, $\alpha=\uparrow,\downarrow$ We fixed $U=U^\prime$ in the calculation of all figures.}
		\label{fig:orb_mag}
	\end{center}
\end{figure}

\section{orbital-moment-induced QAHE with interactions}
Interestingly, once considering onsite Coulomb interaction, we find the instability towards a gapped phase with orbital moment ordering, and the system exhibiting a QAH effect.
Using $d_{xz}$, $d_{yz}$ and $d_{z^2}$ orbitals of Fe atoms, a spinful three-orbital tight-binding model is constructed to capture the DFT band structure (see more details in Appendix~\ref{sup:D}). The Coulomb interaction considered in $d_{xz}$ and $d_{yz}$ orbitals is written as,
\begin{equation}
	H_{int}=U\underset{l}{\sum}n_{l,\uparrow}n_{l,\downarrow}+U^\prime\underset{l<l^\prime,\alpha\beta}{\sum}n_{l,\alpha}n_{l^\prime,\beta},
\end{equation}
where $n_{l,\alpha}$ represents the electron density on orbital $l$ with spin $\alpha$ and $\alpha,\beta=~\uparrow,\downarrow$. We further employ the Hartree-Fock approximation to treat the Coulomb interaction (The results of the LDA+U method are discussed in Appendix~\ref{sup:E}). And the order parameters are defined as
\begin{equation}
	M^{\mu\nu}=\underset{l,l^\prime}{\sum}\underset{\alpha,\beta}{\sum}\tau^{\mu}_{l,l^\prime}\sigma^{\nu}_{\alpha,\beta}\langle d^{\dagger}_{l,\alpha}d_{l^\prime,\beta}\rangle.
\end{equation}
Here, the $d^{\left(\dagger\right)}_{l,\alpha}$ operator annihilates (creates) an electron in orbital $l$ and spin $\alpha$. $\tau^{\mu}$ and $\sigma^{\nu}$ $\left(\mu,\nu=0,x,y,z\right)$ are the identity with three Pauli matrices representing the orbital and spin degree of freedom respectively. We self-consistently investigate the zero-temperature phase diagram, and the results are shown in Fig.~\ref{fig:orb_mag}.

In self-consistent calculations, we exclude the channels for SOC relevant order parameter $M^{yz}$ and the symmetric part of the electron density terms $M^{00}$, which are already considered in the DFT calculation. After considering all other order parameters, we find that the Coulomb interaction between $d_{xz}$ and $d_{yz}$ orbitals intrinsically stimulates orbital-magnetization phases respectively with order parameters $M^{yx}$, $M^{yy}$ and $M^{y0}$. These three orderings break the $TC_{2z}=i\tau_{z}\otimes\sigma_{x}\mathcal{K}$ symmetry and open the gap at the $K$ point. The $M^{yx}$ phase and $M^{yy}$ phase are related by $C_{6z}$ symmetry with relative higher energy, while the $M^{y0}$ phase with orbital moment ordering is the ground state. The gap width exponentially grows when increasing the Coulomb interaction strength $U$ in Fig.\ref{fig:orb_mag}(a). The gapped system then becomes a $C=2$ Chern insulator.

\section{Conclusion}
In conclusion, we have explored electronic structures and the topological property of the FeB$_2$ monolayer.
Without SOC, the FeB$_2$ monolayer has linear band crossing points at $K$ points.
Upon including SOC, they become QBCPs with Berry phase $2\pi$.
Based on effective Hamiltonians, we demonstrate that the QBCPs are not protected by rotational symmetry. 
The appearance of QBCPs (or zero velocity at $K$) is similar to the case in the magic-angle TBG.
FeB$_2$ monolayer is a good platform for studying the instability of spinful QBCPs.
Considering Coulomb interaction in the spinful model of FeB$_2$ with QBCPs, it turns out to be a $C=2$ QAH insulator with orbital moment ordering.

\begin{acknowledgments}
This work was supported by the National Natural Science Foundation of China (Grants No. 11974395, 12188101, U2032204), the Strategic Priority Research Program of Chinese Academy of Sciences (Grant No. XDB33000000), China Postdoctoral Science Foundation funded project (Grant No. 2021M703461), and the Center for Materials Genome.
\end{acknowledgments}


\begin{thebibliography}{43}%
	\makeatletter
	\providecommand \@ifxundefined [1]{%
		\@ifx{#1\undefined}
	}%
	\providecommand \@ifnum [1]{%
		\ifnum #1\expandafter \@firstoftwo
		\else \expandafter \@secondoftwo
		\fi
	}%
	\providecommand \@ifx [1]{%
		\ifx #1\expandafter \@firstoftwo
		\else \expandafter \@secondoftwo
		\fi
	}%
	\providecommand \natexlab [1]{#1}%
	\providecommand \enquote  [1]{``#1''}%
	\providecommand \bibnamefont  [1]{#1}%
	\providecommand \bibfnamefont [1]{#1}%
	\providecommand \citenamefont [1]{#1}%
	\providecommand \href@noop [0]{\@secondoftwo}%
	\providecommand \href [0]{\begingroup \@sanitize@url \@href}%
	\providecommand \@href[1]{\@@startlink{#1}\@@href}%
	\providecommand \@@href[1]{\endgroup#1\@@endlink}%
	\providecommand \@sanitize@url [0]{\catcode `\\12\catcode `\$12\catcode
		`\&12\catcode `\#12\catcode `\^12\catcode `\_12\catcode `\%12\relax}%
	\providecommand \@@startlink[1]{}%
	\providecommand \@@endlink[0]{}%
	\providecommand \url  [0]{\begingroup\@sanitize@url \@url }%
	\providecommand \@url [1]{\endgroup\@href {#1}{\urlprefix }}%
	\providecommand \urlprefix  [0]{URL }%
	\providecommand \Eprint [0]{\href }%
	\providecommand \doibase [0]{http://dx.doi.org/}%
	\providecommand \selectlanguage [0]{\@gobble}%
	\providecommand \bibinfo  [0]{\@secondoftwo}%
	\providecommand \bibfield  [0]{\@secondoftwo}%
	\providecommand \translation [1]{[#1]}%
	\providecommand \BibitemOpen [0]{}%
	\providecommand \bibitemStop [0]{}%
	\providecommand \bibitemNoStop [0]{.\EOS\space}%
	\providecommand \EOS [0]{\spacefactor3000\relax}%
	\providecommand \BibitemShut  [1]{\csname bibitem#1\endcsname}%
	\let\auto@bib@innerbib\@empty
	\bibitem [{\citenamefont {Sun}\ \emph {et~al.}(2009)\citenamefont {Sun},
		\citenamefont {Yao}, \citenamefont {Fradkin},\ and\ \citenamefont
		{Kivelson}}]{quadratic2009}%
	\BibitemOpen
	\bibfield  {author} {\bibinfo {author} {\bibfnamefont {K.}~\bibnamefont
			{Sun}}, \bibinfo {author} {\bibfnamefont {H.}~\bibnamefont {Yao}}, \bibinfo
		{author} {\bibfnamefont {E.}~\bibnamefont {Fradkin}}, \ and\ \bibinfo
		{author} {\bibfnamefont {S.~A.}\ \bibnamefont {Kivelson}},\ }\href {\doibase
		10.1103/PhysRevLett.103.046811} {\bibfield  {journal} {\bibinfo  {journal}
			{Phys. Rev. Lett.}\ }\textbf {\bibinfo {volume} {103}},\ \bibinfo {pages}
		{046811} (\bibinfo {year} {2009})}\BibitemShut {NoStop}%
	\bibitem [{\citenamefont {Hesselmann}\ \emph {et~al.}(2020)\citenamefont
		{Hesselmann}, \citenamefont {Honerkamp}, \citenamefont {Wessel},\ and\
		\citenamefont {Lang}}]{quadratic2020}%
	\BibitemOpen
	\bibfield  {author} {\bibinfo {author} {\bibfnamefont {S.}~\bibnamefont
			{Hesselmann}}, \bibinfo {author} {\bibfnamefont {C.}~\bibnamefont
			{Honerkamp}}, \bibinfo {author} {\bibfnamefont {S.}~\bibnamefont {Wessel}}, \
		and\ \bibinfo {author} {\bibfnamefont {T.~C.}\ \bibnamefont {Lang}},\ }\href
	{\doibase 10.1103/PhysRevB.101.075128} {\bibfield  {journal} {\bibinfo
			{journal} {Phys. Rev. B}\ }\textbf {\bibinfo {volume} {101}},\ \bibinfo
		{pages} {075128} (\bibinfo {year} {2020})}\BibitemShut {NoStop}%
	\bibitem [{\citenamefont {Montambaux}(2012)}]{EPJB2012}%
	\BibitemOpen
	\bibfield  {author} {\bibinfo {author} {\bibfnamefont {G.}~\bibnamefont
			{Montambaux}},\ }\href {\doibase 10.1140/epjb/e2012-30570-7} {\bibfield
		{journal} {\bibinfo  {journal} {The European Physical Journal B}\ }\textbf
		{\bibinfo {volume} {85}},\ \bibinfo {pages} {375} (\bibinfo {year}
		{2012})}\BibitemShut {NoStop}%
	\bibitem [{\citenamefont {Vafek}\ and\ \citenamefont
		{Yang}(2010)}]{quadratic2010}%
	\BibitemOpen
	\bibfield  {author} {\bibinfo {author} {\bibfnamefont {O.}~\bibnamefont
			{Vafek}}\ and\ \bibinfo {author} {\bibfnamefont {K.}~\bibnamefont {Yang}},\
	}\href {\doibase 10.1103/PhysRevB.81.041401} {\bibfield  {journal} {\bibinfo
			{journal} {Phys. Rev. B}\ }\textbf {\bibinfo {volume} {81}},\ \bibinfo
		{pages} {041401} (\bibinfo {year} {2010})}\BibitemShut {NoStop}%
	\bibitem [{\citenamefont {Liang}\ \emph {et~al.}()\citenamefont {Liang},
		\citenamefont {Zhou}, \citenamefont {Yu}, \citenamefont {Wang},\ and\
		\citenamefont {Weng}}]{liang_interaction-driven_2017}%
	\BibitemOpen
	\bibfield  {author} {\bibinfo {author} {\bibfnamefont {Q.-F.}\ \bibnamefont
			{Liang}}, \bibinfo {author} {\bibfnamefont {J.}~\bibnamefont {Zhou}},
		\bibinfo {author} {\bibfnamefont {R.}~\bibnamefont {Yu}}, \bibinfo {author}
		{\bibfnamefont {X.}~\bibnamefont {Wang}}, \ and\ \bibinfo {author}
		{\bibfnamefont {H.}~\bibnamefont {Weng}},\ }\href {\doibase
		10.1103/PhysRevB.96.205412} {\bibfield  {journal} {\bibinfo  {journal} {Phys.
				Rev. B}\ }\textbf {\bibinfo {volume} {96}},\ \bibinfo {pages}
		{205412}}\BibitemShut {NoStop}%
	\bibitem [{\citenamefont {Bernevig}\ \emph {et~al.}(2006)\citenamefont
		{Bernevig}, \citenamefont {Hughes},\ and\ \citenamefont {Zhang}}]{QSH-2006}%
	\BibitemOpen
	\bibfield  {author} {\bibinfo {author} {\bibfnamefont {B.~A.}\ \bibnamefont
			{Bernevig}}, \bibinfo {author} {\bibfnamefont {T.~L.}\ \bibnamefont
			{Hughes}}, \ and\ \bibinfo {author} {\bibfnamefont {S.-C.}\ \bibnamefont
			{Zhang}},\ }\href {\doibase 10.1126/science.1133734} {\bibfield  {journal}
		{\bibinfo  {journal} {Science}\ }\textbf {\bibinfo {volume} {314}},\ \bibinfo
		{pages} {1757} (\bibinfo {year} {2006})}\BibitemShut {NoStop}%
	\bibitem [{\citenamefont {Zhang}\ \emph {et~al.}(2009)\citenamefont {Zhang},
		\citenamefont {Liu}, \citenamefont {Qi}, \citenamefont {Dai}, \citenamefont
		{Fang},\ and\ \citenamefont {Zhang}}]{Bi2Se3-2009}%
	\BibitemOpen
	\bibfield  {author} {\bibinfo {author} {\bibfnamefont {H.}~\bibnamefont
			{Zhang}}, \bibinfo {author} {\bibfnamefont {C.-X.}\ \bibnamefont {Liu}},
		\bibinfo {author} {\bibfnamefont {X.-L.}\ \bibnamefont {Qi}}, \bibinfo
		{author} {\bibfnamefont {X.}~\bibnamefont {Dai}}, \bibinfo {author}
		{\bibfnamefont {Z.}~\bibnamefont {Fang}}, \ and\ \bibinfo {author}
		{\bibfnamefont {S.-C.}\ \bibnamefont {Zhang}},\ }\href {\doibase
		10.1038/nphys1270} {\bibfield  {journal} {\bibinfo  {journal} {Nat. Phys.}\
		}\textbf {\bibinfo {volume} {5}},\ \bibinfo {pages} {438} (\bibinfo {year}
		{2009})}\BibitemShut {NoStop}%
	\bibitem [{\citenamefont {Nie}\ \emph {et~al.}(2018)\citenamefont {Nie},
		\citenamefont {Xing}, \citenamefont {Jin}, \citenamefont {Xie}, \citenamefont
		{Wang},\ and\ \citenamefont {Prinz}}]{TaSe3_2018}%
	\BibitemOpen
	\bibfield  {author} {\bibinfo {author} {\bibfnamefont {S.}~\bibnamefont
			{Nie}}, \bibinfo {author} {\bibfnamefont {L.}~\bibnamefont {Xing}}, \bibinfo
		{author} {\bibfnamefont {R.}~\bibnamefont {Jin}}, \bibinfo {author}
		{\bibfnamefont {W.}~\bibnamefont {Xie}}, \bibinfo {author} {\bibfnamefont
			{Z.}~\bibnamefont {Wang}}, \ and\ \bibinfo {author} {\bibfnamefont {F.~B.}\
			\bibnamefont {Prinz}},\ }\href {\doibase 10.1103/PhysRevB.98.125143}
	{\bibfield  {journal} {\bibinfo  {journal} {Phys. Rev. B}\ }\textbf {\bibinfo
			{volume} {98}},\ \bibinfo {pages} {125143} (\bibinfo {year}
		{2018})}\BibitemShut {NoStop}%
	\bibitem [{\citenamefont {Wan}\ \emph {et~al.}(2011)\citenamefont {Wan},
		\citenamefont {Turner}, \citenamefont {Vishwanath},\ and\ \citenamefont
		{Savrasov}}]{Weyl-2009}%
	\BibitemOpen
	\bibfield  {author} {\bibinfo {author} {\bibfnamefont {X.}~\bibnamefont
			{Wan}}, \bibinfo {author} {\bibfnamefont {A.~M.}\ \bibnamefont {Turner}},
		\bibinfo {author} {\bibfnamefont {A.}~\bibnamefont {Vishwanath}}, \ and\
		\bibinfo {author} {\bibfnamefont {S.~Y.}\ \bibnamefont {Savrasov}},\ }\href
	{\doibase 10.1103/PhysRevB.83.205101} {\bibfield  {journal} {\bibinfo
			{journal} {Phys. Rev. B}\ }\textbf {\bibinfo {volume} {83}},\ \bibinfo
		{pages} {205101} (\bibinfo {year} {2011})}\BibitemShut {NoStop}%
	\bibitem [{\citenamefont {Weng}\ \emph {et~al.}(2015)\citenamefont {Weng},
		\citenamefont {Fang}, \citenamefont {Fang}, \citenamefont {Bernevig},\ and\
		\citenamefont {Dai}}]{TaAs-2015-Weng}%
	\BibitemOpen
	\bibfield  {author} {\bibinfo {author} {\bibfnamefont {H.}~\bibnamefont
			{Weng}}, \bibinfo {author} {\bibfnamefont {C.}~\bibnamefont {Fang}}, \bibinfo
		{author} {\bibfnamefont {Z.}~\bibnamefont {Fang}}, \bibinfo {author}
		{\bibfnamefont {B.~A.}\ \bibnamefont {Bernevig}}, \ and\ \bibinfo {author}
		{\bibfnamefont {X.}~\bibnamefont {Dai}},\ }\href {\doibase
		10.1103/PhysRevX.5.011029} {\bibfield  {journal} {\bibinfo  {journal} {Phys.
				Rev. X}\ }\textbf {\bibinfo {volume} {5}},\ \bibinfo {pages} {011029}
		(\bibinfo {year} {2015})}\BibitemShut {NoStop}%
	\bibitem [{\citenamefont {Xu}\ \emph {et~al.}(2015)\citenamefont {Xu},
		\citenamefont {Belopolski}, \citenamefont {Alidoust}, \citenamefont
		{Neupane}, \citenamefont {Bian}, \citenamefont {Zhang}, \citenamefont
		{Sankar}, \citenamefont {Chang}, \citenamefont {Yuan}, \citenamefont {Lee},
		\citenamefont {Huang}, \citenamefont {Zheng}, \citenamefont {Ma},
		\citenamefont {Sanchez}, \citenamefont {Wang}, \citenamefont {Bansil},
		\citenamefont {Chou}, \citenamefont {Shibayev}, \citenamefont {Lin},
		\citenamefont {Jia},\ and\ \citenamefont {Hasan}}]{TaAs-2015-Xu}%
	\BibitemOpen
	\bibfield  {author} {\bibinfo {author} {\bibfnamefont {S.-Y.}\ \bibnamefont
			{Xu}}, \bibinfo {author} {\bibfnamefont {I.}~\bibnamefont {Belopolski}},
		\bibinfo {author} {\bibfnamefont {N.}~\bibnamefont {Alidoust}}, \bibinfo
		{author} {\bibfnamefont {M.}~\bibnamefont {Neupane}}, \bibinfo {author}
		{\bibfnamefont {G.}~\bibnamefont {Bian}}, \bibinfo {author} {\bibfnamefont
			{C.}~\bibnamefont {Zhang}}, \bibinfo {author} {\bibfnamefont
			{R.}~\bibnamefont {Sankar}}, \bibinfo {author} {\bibfnamefont
			{G.}~\bibnamefont {Chang}}, \bibinfo {author} {\bibfnamefont
			{Z.}~\bibnamefont {Yuan}}, \bibinfo {author} {\bibfnamefont {C.-C.}\
			\bibnamefont {Lee}}, \bibinfo {author} {\bibfnamefont {S.-M.}\ \bibnamefont
			{Huang}}, \bibinfo {author} {\bibfnamefont {H.}~\bibnamefont {Zheng}},
		\bibinfo {author} {\bibfnamefont {J.}~\bibnamefont {Ma}}, \bibinfo {author}
		{\bibfnamefont {D.~S.}\ \bibnamefont {Sanchez}}, \bibinfo {author}
		{\bibfnamefont {B.}~\bibnamefont {Wang}}, \bibinfo {author} {\bibfnamefont
			{A.}~\bibnamefont {Bansil}}, \bibinfo {author} {\bibfnamefont
			{F.}~\bibnamefont {Chou}}, \bibinfo {author} {\bibfnamefont {P.~P.}\
			\bibnamefont {Shibayev}}, \bibinfo {author} {\bibfnamefont {H.}~\bibnamefont
			{Lin}}, \bibinfo {author} {\bibfnamefont {S.}~\bibnamefont {Jia}}, \ and\
		\bibinfo {author} {\bibfnamefont {M.~Z.}\ \bibnamefont {Hasan}},\ }\href
	{\doibase 10.1126/science.aaa9297} {\bibfield  {journal} {\bibinfo  {journal}
			{Science}\ }\textbf {\bibinfo {volume} {349}},\ \bibinfo {pages} {613}
		(\bibinfo {year} {2015})}\BibitemShut {NoStop}%
	\bibitem [{\citenamefont {Wang}\ \emph {et~al.}(2016)\citenamefont {Wang},
		\citenamefont {Vergniory}, \citenamefont {Kushwaha}, \citenamefont
		{Hirschberger}, \citenamefont {Chulkov}, \citenamefont {Ernst}, \citenamefont
		{Ong}, \citenamefont {Cava},\ and\ \citenamefont {Bernevig}}]{ZrCo2Sn2016}%
	\BibitemOpen
	\bibfield  {author} {\bibinfo {author} {\bibfnamefont {Z.}~\bibnamefont
			{Wang}}, \bibinfo {author} {\bibfnamefont {M.~G.}\ \bibnamefont {Vergniory}},
		\bibinfo {author} {\bibfnamefont {S.}~\bibnamefont {Kushwaha}}, \bibinfo
		{author} {\bibfnamefont {M.}~\bibnamefont {Hirschberger}}, \bibinfo {author}
		{\bibfnamefont {E.~V.}\ \bibnamefont {Chulkov}}, \bibinfo {author}
		{\bibfnamefont {A.}~\bibnamefont {Ernst}}, \bibinfo {author} {\bibfnamefont
			{N.~P.}\ \bibnamefont {Ong}}, \bibinfo {author} {\bibfnamefont {R.~J.}\
			\bibnamefont {Cava}}, \ and\ \bibinfo {author} {\bibfnamefont {B.~A.}\
			\bibnamefont {Bernevig}},\ }\href {\doibase 10.1103/PhysRevLett.117.236401}
	{\bibfield  {journal} {\bibinfo  {journal} {Phys. Rev. Lett.}\ }\textbf
		{\bibinfo {volume} {117}},\ \bibinfo {pages} {236401} (\bibinfo {year}
		{2016})}\BibitemShut {NoStop}%
	\bibitem [{\citenamefont {Thouless}\ \emph {et~al.}(1982)\citenamefont
		{Thouless}, \citenamefont {Kohmoto}, \citenamefont {Nightingale},\ and\
		\citenamefont {Dennijs}}]{TKNN}%
	\BibitemOpen
	\bibfield  {author} {\bibinfo {author} {\bibfnamefont {D.~J.}\ \bibnamefont
			{Thouless}}, \bibinfo {author} {\bibfnamefont {M.}~\bibnamefont {Kohmoto}},
		\bibinfo {author} {\bibfnamefont {M.~P.}\ \bibnamefont {Nightingale}}, \ and\
		\bibinfo {author} {\bibfnamefont {M.}~\bibnamefont {Dennijs}},\ }\href
	{\doibase 10.1103/PhysRevLett.49.405} {\bibfield  {journal} {\bibinfo
			{journal} {Phys. Rev. Lett.}\ }\textbf {\bibinfo {volume} {49}},\ \bibinfo
		{pages} {405} (\bibinfo {year} {1982})}\BibitemShut {NoStop}%
	\bibitem [{\citenamefont {Yu}\ \emph {et~al.}(2010)\citenamefont {Yu},
		\citenamefont {Zhang}, \citenamefont {Zhang}, \citenamefont {Zhang},
		\citenamefont {Dai},\ and\ \citenamefont {Fang}}]{Yu2010}%
	\BibitemOpen
	\bibfield  {author} {\bibinfo {author} {\bibfnamefont {R.}~\bibnamefont
			{Yu}}, \bibinfo {author} {\bibfnamefont {W.}~\bibnamefont {Zhang}}, \bibinfo
		{author} {\bibfnamefont {H.-J.}\ \bibnamefont {Zhang}}, \bibinfo {author}
		{\bibfnamefont {S.-C.}\ \bibnamefont {Zhang}}, \bibinfo {author}
		{\bibfnamefont {X.}~\bibnamefont {Dai}}, \ and\ \bibinfo {author}
		{\bibfnamefont {Z.}~\bibnamefont {Fang}},\ }\href {\doibase
		10.1126/science.1187485} {\bibfield  {journal} {\bibinfo  {journal}
			{Science}\ }\textbf {\bibinfo {volume} {329}},\ \bibinfo {pages} {61}
		(\bibinfo {year} {2010})}\BibitemShut {NoStop}%
	\bibitem [{\citenamefont {Qiao}\ \emph {et~al.}(2010)\citenamefont {Qiao},
		\citenamefont {Yang}, \citenamefont {Feng}, \citenamefont {Tse},
		\citenamefont {Ding}, \citenamefont {Yao}, \citenamefont {Wang},\ and\
		\citenamefont {Niu}}]{Qiao2010}%
	\BibitemOpen
	\bibfield  {author} {\bibinfo {author} {\bibfnamefont {Z.}~\bibnamefont
			{Qiao}}, \bibinfo {author} {\bibfnamefont {S.~A.}\ \bibnamefont {Yang}},
		\bibinfo {author} {\bibfnamefont {W.}~\bibnamefont {Feng}}, \bibinfo {author}
		{\bibfnamefont {W.-K.}\ \bibnamefont {Tse}}, \bibinfo {author} {\bibfnamefont
			{J.}~\bibnamefont {Ding}}, \bibinfo {author} {\bibfnamefont {Y.}~\bibnamefont
			{Yao}}, \bibinfo {author} {\bibfnamefont {J.}~\bibnamefont {Wang}}, \ and\
		\bibinfo {author} {\bibfnamefont {Q.}~\bibnamefont {Niu}},\ }\href {\doibase
		10.1103/PhysRevB.82.161414} {\bibfield  {journal} {\bibinfo  {journal} {Phys.
				Rev. B}\ }\textbf {\bibinfo {volume} {82}},\ \bibinfo {pages} {161414}
		(\bibinfo {year} {2010})}\BibitemShut {NoStop}%
	\bibitem [{\citenamefont {Garrity}\ and\ \citenamefont
		{Vanderbilt}(2013)}]{PhysRevLett.110.116802}%
	\BibitemOpen
	\bibfield  {author} {\bibinfo {author} {\bibfnamefont {K.~F.}\ \bibnamefont
			{Garrity}}\ and\ \bibinfo {author} {\bibfnamefont {D.}~\bibnamefont
			{Vanderbilt}},\ }\href {\doibase 10.1103/PhysRevLett.110.116802} {\bibfield
		{journal} {\bibinfo  {journal} {Phys. Rev. Lett.}\ }\textbf {\bibinfo
			{volume} {110}},\ \bibinfo {pages} {116802} (\bibinfo {year}
		{2013})}\BibitemShut {NoStop}%
	\bibitem [{\citenamefont {Xue}\ \emph {et~al.}(2018)\citenamefont {Xue},
		\citenamefont {Zhang}, \citenamefont {Zhao}, \citenamefont {Wei},\ and\
		\citenamefont {Yang}}]{Xue2018}%
	\BibitemOpen
	\bibfield  {author} {\bibinfo {author} {\bibfnamefont {Y.}~\bibnamefont
			{Xue}}, \bibinfo {author} {\bibfnamefont {J.~Y.}\ \bibnamefont {Zhang}},
		\bibinfo {author} {\bibfnamefont {B.}~\bibnamefont {Zhao}}, \bibinfo {author}
		{\bibfnamefont {X.~Y.}\ \bibnamefont {Wei}}, \ and\ \bibinfo {author}
		{\bibfnamefont {Z.~Q.}\ \bibnamefont {Yang}},\ }\href {\doibase
		10.1039/c8nr00201k} {\bibfield  {journal} {\bibinfo  {journal} {Nanoscale}\
		}\textbf {\bibinfo {volume} {10}},\ \bibinfo {pages} {8569} (\bibinfo {year}
		{2018})}\BibitemShut {NoStop}%
	\bibitem [{\citenamefont {Wang}\ \emph {et~al.}(2013)\citenamefont {Wang},
		\citenamefont {Liu},\ and\ \citenamefont {Liu}}]{PhysRevLett.110.196801}%
	\BibitemOpen
	\bibfield  {author} {\bibinfo {author} {\bibfnamefont {Z.~F.}\ \bibnamefont
			{Wang}}, \bibinfo {author} {\bibfnamefont {Z.}~\bibnamefont {Liu}}, \ and\
		\bibinfo {author} {\bibfnamefont {F.}~\bibnamefont {Liu}},\ }\href {\doibase
		10.1103/PhysRevLett.110.196801} {\bibfield  {journal} {\bibinfo  {journal}
			{Phys. Rev. Lett.}\ }\textbf {\bibinfo {volume} {110}},\ \bibinfo {pages}
		{196801} (\bibinfo {year} {2013})}\BibitemShut {NoStop}%
	\bibitem [{\citenamefont {Nie}\ \emph {et~al.}(2020)\citenamefont {Nie},
		\citenamefont {Sun}, \citenamefont {Prinz}, \citenamefont {Wang},
		\citenamefont {Weng}, \citenamefont {Fang},\ and\ \citenamefont
		{Dai}}]{EuB6_2020}%
	\BibitemOpen
	\bibfield  {author} {\bibinfo {author} {\bibfnamefont {S.}~\bibnamefont
			{Nie}}, \bibinfo {author} {\bibfnamefont {Y.}~\bibnamefont {Sun}}, \bibinfo
		{author} {\bibfnamefont {F.~B.}\ \bibnamefont {Prinz}}, \bibinfo {author}
		{\bibfnamefont {Z.}~\bibnamefont {Wang}}, \bibinfo {author} {\bibfnamefont
			{H.}~\bibnamefont {Weng}}, \bibinfo {author} {\bibfnamefont {Z.}~\bibnamefont
			{Fang}}, \ and\ \bibinfo {author} {\bibfnamefont {X.}~\bibnamefont {Dai}},\
	}\href {\doibase 10.1103/PhysRevLett.124.076403} {\bibfield  {journal}
		{\bibinfo  {journal} {Phys. Rev. Lett.}\ }\textbf {\bibinfo {volume} {124}},\
		\bibinfo {pages} {076403} (\bibinfo {year} {2020})}\BibitemShut {NoStop}%
	\bibitem [{\citenamefont {Chang}\ \emph {et~al.}(2013)\citenamefont {Chang},
		\citenamefont {Zhang}, \citenamefont {Feng}, \citenamefont {Shen},
		\citenamefont {Zhang}, \citenamefont {Guo}, \citenamefont {Li}, \citenamefont
		{Ou}, \citenamefont {Wei}, \citenamefont {Wang}, \citenamefont {Ji},
		\citenamefont {Feng}, \citenamefont {Ji}, \citenamefont {Chen}, \citenamefont
		{Jia}, \citenamefont {Dai}, \citenamefont {Fang}, \citenamefont {Zhang},
		\citenamefont {He}, \citenamefont {Wang}, \citenamefont {Lu}, \citenamefont
		{Ma},\ and\ \citenamefont {Xue}}]{Chang2013}%
	\BibitemOpen
	\bibfield  {author} {\bibinfo {author} {\bibfnamefont {C.-Z.}\ \bibnamefont
			{Chang}}, \bibinfo {author} {\bibfnamefont {J.}~\bibnamefont {Zhang}},
		\bibinfo {author} {\bibfnamefont {X.}~\bibnamefont {Feng}}, \bibinfo {author}
		{\bibfnamefont {J.}~\bibnamefont {Shen}}, \bibinfo {author} {\bibfnamefont
			{Z.}~\bibnamefont {Zhang}}, \bibinfo {author} {\bibfnamefont
			{M.}~\bibnamefont {Guo}}, \bibinfo {author} {\bibfnamefont {K.}~\bibnamefont
			{Li}}, \bibinfo {author} {\bibfnamefont {Y.}~\bibnamefont {Ou}}, \bibinfo
		{author} {\bibfnamefont {P.}~\bibnamefont {Wei}}, \bibinfo {author}
		{\bibfnamefont {L.-L.}\ \bibnamefont {Wang}}, \bibinfo {author}
		{\bibfnamefont {Z.-Q.}\ \bibnamefont {Ji}}, \bibinfo {author} {\bibfnamefont
			{Y.}~\bibnamefont {Feng}}, \bibinfo {author} {\bibfnamefont {S.}~\bibnamefont
			{Ji}}, \bibinfo {author} {\bibfnamefont {X.}~\bibnamefont {Chen}}, \bibinfo
		{author} {\bibfnamefont {J.}~\bibnamefont {Jia}}, \bibinfo {author}
		{\bibfnamefont {X.}~\bibnamefont {Dai}}, \bibinfo {author} {\bibfnamefont
			{Z.}~\bibnamefont {Fang}}, \bibinfo {author} {\bibfnamefont {S.-C.}\
			\bibnamefont {Zhang}}, \bibinfo {author} {\bibfnamefont {K.}~\bibnamefont
			{He}}, \bibinfo {author} {\bibfnamefont {Y.}~\bibnamefont {Wang}}, \bibinfo
		{author} {\bibfnamefont {L.}~\bibnamefont {Lu}}, \bibinfo {author}
		{\bibfnamefont {X.-C.}\ \bibnamefont {Ma}}, \ and\ \bibinfo {author}
		{\bibfnamefont {Q.-K.}\ \bibnamefont {Xue}},\ }\href {\doibase
		10.1126/science.1234414} {\bibfield  {journal} {\bibinfo  {journal}
			{Science}\ }\textbf {\bibinfo {volume} {340}},\ \bibinfo {pages} {167}
		(\bibinfo {year} {2013})}\BibitemShut {NoStop}%
	\bibitem [{\citenamefont {Chang}\ \emph {et~al.}(2015)\citenamefont {Chang},
		\citenamefont {Zhao}, \citenamefont {Kim}, \citenamefont {Zhang},
		\citenamefont {Assaf}, \citenamefont {Heiman}, \citenamefont {Zhang},
		\citenamefont {Liu}, \citenamefont {Chan},\ and\ \citenamefont
		{Moodera}}]{Chang2015}%
	\BibitemOpen
	\bibfield  {author} {\bibinfo {author} {\bibfnamefont {C.-Z.}\ \bibnamefont
			{Chang}}, \bibinfo {author} {\bibfnamefont {W.}~\bibnamefont {Zhao}},
		\bibinfo {author} {\bibfnamefont {D.~Y.}\ \bibnamefont {Kim}}, \bibinfo
		{author} {\bibfnamefont {H.}~\bibnamefont {Zhang}}, \bibinfo {author}
		{\bibfnamefont {B.~A.}\ \bibnamefont {Assaf}}, \bibinfo {author}
		{\bibfnamefont {D.}~\bibnamefont {Heiman}}, \bibinfo {author} {\bibfnamefont
			{S.-C.}\ \bibnamefont {Zhang}}, \bibinfo {author} {\bibfnamefont
			{C.}~\bibnamefont {Liu}}, \bibinfo {author} {\bibfnamefont {M.~H.~W.}\
			\bibnamefont {Chan}}, \ and\ \bibinfo {author} {\bibfnamefont {J.~S.}\
			\bibnamefont {Moodera}},\ }\href {\doibase 10.1038/nmat4204} {\bibfield
		{journal} {\bibinfo  {journal} {Nat. Mater.}\ }\textbf {\bibinfo {volume}
			{14}},\ \bibinfo {pages} {473} (\bibinfo {year} {2015})}\BibitemShut
	{NoStop}%
	\bibitem [{\citenamefont {Deng}\ \emph {et~al.}(2020)\citenamefont {Deng},
		\citenamefont {Yu}, \citenamefont {Shi}, \citenamefont {Guo}, \citenamefont
		{Xu}, \citenamefont {Wang}, \citenamefont {Chen},\ and\ \citenamefont
		{Zhang}}]{QAH-MnBi2Te4}%
	\BibitemOpen
	\bibfield  {author} {\bibinfo {author} {\bibfnamefont {Y.}~\bibnamefont
			{Deng}}, \bibinfo {author} {\bibfnamefont {Y.}~\bibnamefont {Yu}}, \bibinfo
		{author} {\bibfnamefont {M.~Z.}\ \bibnamefont {Shi}}, \bibinfo {author}
		{\bibfnamefont {Z.}~\bibnamefont {Guo}}, \bibinfo {author} {\bibfnamefont
			{Z.}~\bibnamefont {Xu}}, \bibinfo {author} {\bibfnamefont {J.}~\bibnamefont
			{Wang}}, \bibinfo {author} {\bibfnamefont {X.~H.}\ \bibnamefont {Chen}}, \
		and\ \bibinfo {author} {\bibfnamefont {Y.}~\bibnamefont {Zhang}},\ }\href
	{\doibase 10.1126/science.aax8156} {\bibfield  {journal} {\bibinfo  {journal}
			{Science}\ }\textbf {\bibinfo {volume} {367}},\ \bibinfo {pages} {895}
		(\bibinfo {year} {2020})}\BibitemShut {NoStop}%
	\bibitem [{\citenamefont {Serlin}\ \emph {et~al.}(2020)\citenamefont {Serlin},
		\citenamefont {Tschirhart}, \citenamefont {Polshyn}, \citenamefont {Zhang},
		\citenamefont {Zhu}, \citenamefont {Watanabe}, \citenamefont {Taniguchi},
		\citenamefont {Balents},\ and\ \citenamefont {Young}}]{QAH-TBG}%
	\BibitemOpen
	\bibfield  {author} {\bibinfo {author} {\bibfnamefont {M.}~\bibnamefont
			{Serlin}}, \bibinfo {author} {\bibfnamefont {C.~L.}\ \bibnamefont
			{Tschirhart}}, \bibinfo {author} {\bibfnamefont {H.}~\bibnamefont {Polshyn}},
		\bibinfo {author} {\bibfnamefont {Y.}~\bibnamefont {Zhang}}, \bibinfo
		{author} {\bibfnamefont {J.}~\bibnamefont {Zhu}}, \bibinfo {author}
		{\bibfnamefont {K.}~\bibnamefont {Watanabe}}, \bibinfo {author}
		{\bibfnamefont {T.}~\bibnamefont {Taniguchi}}, \bibinfo {author}
		{\bibfnamefont {L.}~\bibnamefont {Balents}}, \ and\ \bibinfo {author}
		{\bibfnamefont {A.~F.}\ \bibnamefont {Young}},\ }\href {\doibase
		10.1126/science.aay5533} {\bibfield  {journal} {\bibinfo  {journal}
			{Science}\ }\textbf {\bibinfo {volume} {367}},\ \bibinfo {pages} {900}
		(\bibinfo {year} {2020})}\BibitemShut {NoStop}%
	\bibitem [{\citenamefont {Zhang}\ \emph {et~al.}(2014)\citenamefont {Zhang},
		\citenamefont {Wang}, \citenamefont {Du}, \citenamefont {Gao},\ and\
		\citenamefont {Liu}}]{TiB2-2014}%
	\BibitemOpen
	\bibfield  {author} {\bibinfo {author} {\bibfnamefont {L.~Z.}\ \bibnamefont
			{Zhang}}, \bibinfo {author} {\bibfnamefont {Z.~F.}\ \bibnamefont {Wang}},
		\bibinfo {author} {\bibfnamefont {S.~X.}\ \bibnamefont {Du}}, \bibinfo
		{author} {\bibfnamefont {H.~J.}\ \bibnamefont {Gao}}, \ and\ \bibinfo
		{author} {\bibfnamefont {F.}~\bibnamefont {Liu}},\ }\href {\doibase
		10.1103/PhysRevB.90.161402} {\bibfield  {journal} {\bibinfo  {journal} {Phys.
				Rev. B}\ }\textbf {\bibinfo {volume} {90}},\ \bibinfo {pages} {161402}
		(\bibinfo {year} {2014})}\BibitemShut {NoStop}%
	\bibitem [{\citenamefont {Zhang}\ \emph {et~al.}(2016)\citenamefont {Zhang},
		\citenamefont {Li}, \citenamefont {Hou}, \citenamefont {Du},\ and\
		\citenamefont {Chen}}]{FeB2-2016}%
	\BibitemOpen
	\bibfield  {author} {\bibinfo {author} {\bibfnamefont {H.}~\bibnamefont
			{Zhang}}, \bibinfo {author} {\bibfnamefont {Y.}~\bibnamefont {Li}}, \bibinfo
		{author} {\bibfnamefont {J.}~\bibnamefont {Hou}}, \bibinfo {author}
		{\bibfnamefont {A.}~\bibnamefont {Du}}, \ and\ \bibinfo {author}
		{\bibfnamefont {Z.}~\bibnamefont {Chen}},\ }\href {\doibase
		10.1021/acs.nanolett.6b02335} {\bibfield  {journal} {\bibinfo  {journal}
			{Nano Lett.}\ }\textbf {\bibinfo {volume} {16}},\ \bibinfo {pages} {6124}
		(\bibinfo {year} {2016})}\BibitemShut {NoStop}%
	\bibitem [{\citenamefont {Liu}\ \emph {et~al.}(2019)\citenamefont {Liu},
		\citenamefont {Wang}, \citenamefont {Cui}, \citenamefont {Yang},
		\citenamefont {Jin},\ and\ \citenamefont {Xiong}}]{HfB2-2019}%
	\BibitemOpen
	\bibfield  {author} {\bibinfo {author} {\bibfnamefont {Z.}~\bibnamefont
			{Liu}}, \bibinfo {author} {\bibfnamefont {P.}~\bibnamefont {Wang}}, \bibinfo
		{author} {\bibfnamefont {Q.}~\bibnamefont {Cui}}, \bibinfo {author}
		{\bibfnamefont {G.}~\bibnamefont {Yang}}, \bibinfo {author} {\bibfnamefont
			{S.}~\bibnamefont {Jin}}, \ and\ \bibinfo {author} {\bibfnamefont
			{K.}~\bibnamefont {Xiong}},\ }\href {\doibase 10.1039/c8ra08291j} {\bibfield
		{journal} {\bibinfo  {journal} {Rsc Advances}\ }\textbf {\bibinfo {volume}
			{9}},\ \bibinfo {pages} {2740} (\bibinfo {year} {2019})}\BibitemShut
	{NoStop}%
	\bibitem [{\citenamefont {Voroshin}\ \emph {et~al.}(1970)\citenamefont
		{Voroshin}, \citenamefont {Lyakhovich}, \citenamefont {Panich},\ and\
		\citenamefont {Protasevich}}]{FeB2-1970}%
	\BibitemOpen
	\bibfield  {author} {\bibinfo {author} {\bibfnamefont {L.}~\bibnamefont
			{Voroshin}}, \bibinfo {author} {\bibfnamefont {L.}~\bibnamefont
			{Lyakhovich}}, \bibinfo {author} {\bibfnamefont {G.}~\bibnamefont {Panich}},
		\ and\ \bibinfo {author} {\bibfnamefont {G.}~\bibnamefont {Protasevich}},\
	}\href {\doibase 10.1007/BF00652720} {\bibfield  {journal} {\bibinfo
			{journal} {Met. Sci. Heat Treat.}\ }\textbf {\bibinfo {volume} {12}},\
		\bibinfo {pages} {732} (\bibinfo {year} {1970})}\BibitemShut {NoStop}%
	\bibitem [{\citenamefont {Yang}\ \emph {et~al.}(2018)\citenamefont {Yang},
		\citenamefont {Dai}, \citenamefont {Zhao},\ and\ \citenamefont
		{Meng}}]{FeB2-2018}%
	\BibitemOpen
	\bibfield  {author} {\bibinfo {author} {\bibfnamefont {X.}~\bibnamefont
			{Yang}}, \bibinfo {author} {\bibfnamefont {Z.}~\bibnamefont {Dai}}, \bibinfo
		{author} {\bibfnamefont {Y.}~\bibnamefont {Zhao}}, \ and\ \bibinfo {author}
		{\bibfnamefont {S.}~\bibnamefont {Meng}},\ }\href {\doibase
		10.1016/j.commatsci.2018.01.050} {\bibfield  {journal} {\bibinfo  {journal}
			{Comp. Mater. Sci.}\ }\textbf {\bibinfo {volume} {147}},\ \bibinfo {pages}
		{132} (\bibinfo {year} {2018})}\BibitemShut {NoStop}%
	\bibitem [{\citenamefont {Ahmadi}\ \emph {et~al.}(2019)\citenamefont {Ahmadi},
		\citenamefont {Masoudi}, \citenamefont {Taghizade}, \citenamefont {Jafari},\
		and\ \citenamefont {Faghihnasiri}}]{FeB2-2019}%
	\BibitemOpen
	\bibfield  {author} {\bibinfo {author} {\bibfnamefont {A.}~\bibnamefont
			{Ahmadi}}, \bibinfo {author} {\bibfnamefont {M.}~\bibnamefont {Masoudi}},
		\bibinfo {author} {\bibfnamefont {N.}~\bibnamefont {Taghizade}}, \bibinfo
		{author} {\bibfnamefont {H.}~\bibnamefont {Jafari}}, \ and\ \bibinfo {author}
		{\bibfnamefont {M.}~\bibnamefont {Faghihnasiri}},\ }\href {\doibase
		10.1016/j.physe.2019.03.011} {\bibfield  {journal} {\bibinfo  {journal}
			{Phys. E Low-dimens. Syst. Nanostruct,}\ }\textbf {\bibinfo {volume} {112}},\
		\bibinfo {pages} {71} (\bibinfo {year} {2019})}\BibitemShut {NoStop}%
	\bibitem [{\citenamefont {Blochl}(1994)}]{PAW1994}%
	\BibitemOpen
	\bibfield  {author} {\bibinfo {author} {\bibfnamefont {P.~E.}\ \bibnamefont
			{Blochl}},\ }\href {\doibase 10.1103/PhysRevB.50.17953} {\bibfield  {journal}
		{\bibinfo  {journal} {Phys. Rev. B}\ }\textbf {\bibinfo {volume} {50}},\
		\bibinfo {pages} {17953} (\bibinfo {year} {1994})}\BibitemShut {NoStop}%
	\bibitem [{\citenamefont {Kresse}\ and\ \citenamefont
		{Joubert}(1999)}]{PAW1999}%
	\BibitemOpen
	\bibfield  {author} {\bibinfo {author} {\bibfnamefont {G.}~\bibnamefont
			{Kresse}}\ and\ \bibinfo {author} {\bibfnamefont {D.}~\bibnamefont
			{Joubert}},\ }\href {\doibase 10.1103/PhysRevB.59.1758} {\bibfield  {journal}
		{\bibinfo  {journal} {Phys. Rev. B}\ }\textbf {\bibinfo {volume} {59}},\
		\bibinfo {pages} {1758} (\bibinfo {year} {1999})}\BibitemShut {NoStop}%
	\bibitem [{\citenamefont {Kresse}\ and\ \citenamefont
		{Furthmuller}(1996{\natexlab{a}})}]{VASP1}%
	\BibitemOpen
	\bibfield  {author} {\bibinfo {author} {\bibfnamefont {G.}~\bibnamefont
			{Kresse}}\ and\ \bibinfo {author} {\bibfnamefont {J.}~\bibnamefont
			{Furthmuller}},\ }\href {\doibase 10.1016/0927-0256(96)00008-0} {\bibfield
		{journal} {\bibinfo  {journal} {Comp. Mater. Sci.}\ }\textbf {\bibinfo
			{volume} {6}},\ \bibinfo {pages} {15} (\bibinfo {year}
		{1996}{\natexlab{a}})}\BibitemShut {NoStop}%
	\bibitem [{\citenamefont {Kresse}\ and\ \citenamefont
		{Furthmuller}(1996{\natexlab{b}})}]{VASP2}%
	\BibitemOpen
	\bibfield  {author} {\bibinfo {author} {\bibfnamefont {G.}~\bibnamefont
			{Kresse}}\ and\ \bibinfo {author} {\bibfnamefont {J.}~\bibnamefont
			{Furthmuller}},\ }\href {\doibase 10.1103/PhysRevB.54.11169} {\bibfield
		{journal} {\bibinfo  {journal} {Phys. Rev. B}\ }\textbf {\bibinfo {volume}
			{54}},\ \bibinfo {pages} {11169} (\bibinfo {year}
		{1996}{\natexlab{b}})}\BibitemShut {NoStop}%
	\bibitem [{\citenamefont {Perdew}\ \emph {et~al.}(1996)\citenamefont {Perdew},
		\citenamefont {Burke},\ and\ \citenamefont {Ernzerhof}}]{PBE}%
	\BibitemOpen
	\bibfield  {author} {\bibinfo {author} {\bibfnamefont {J.~P.}\ \bibnamefont
			{Perdew}}, \bibinfo {author} {\bibfnamefont {K.}~\bibnamefont {Burke}}, \
		and\ \bibinfo {author} {\bibfnamefont {M.}~\bibnamefont {Ernzerhof}},\ }\href
	{\doibase 10.1103/PhysRevLett.77.3865} {\bibfield  {journal} {\bibinfo
			{journal} {Phys. Rev. Lett.}\ }\textbf {\bibinfo {volume} {77}},\ \bibinfo
		{pages} {3865} (\bibinfo {year} {1996})}\BibitemShut {NoStop}%
	\bibitem [{\citenamefont {Dudarev}\ \emph {et~al.}(1998)\citenamefont
		{Dudarev}, \citenamefont {Botton}, \citenamefont {Savrasov}, \citenamefont
		{Humphreys},\ and\ \citenamefont {Sutton}}]{ldau}%
	\BibitemOpen
	\bibfield  {author} {\bibinfo {author} {\bibfnamefont {S.~L.}\ \bibnamefont
			{Dudarev}}, \bibinfo {author} {\bibfnamefont {G.~A.}\ \bibnamefont {Botton}},
		\bibinfo {author} {\bibfnamefont {S.~Y.}\ \bibnamefont {Savrasov}}, \bibinfo
		{author} {\bibfnamefont {C.~J.}\ \bibnamefont {Humphreys}}, \ and\ \bibinfo
		{author} {\bibfnamefont {A.~P.}\ \bibnamefont {Sutton}},\ }\href {\doibase
		10.1103/PhysRevB.57.1505} {\bibfield  {journal} {\bibinfo  {journal} {Phys.
				Rev. B}\ }\textbf {\bibinfo {volume} {57}},\ \bibinfo {pages} {1505}
		(\bibinfo {year} {1998})}\BibitemShut {NoStop}%
	\bibitem [{\citenamefont {Gao}\ \emph {et~al.}(2021)\citenamefont {Gao},
		\citenamefont {Wu}, \citenamefont {Persson},\ and\ \citenamefont
		{Wang}}]{Irvsp}%
	\BibitemOpen
	\bibfield  {author} {\bibinfo {author} {\bibfnamefont {J.~C.}\ \bibnamefont
			{Gao}}, \bibinfo {author} {\bibfnamefont {Q.~S.}\ \bibnamefont {Wu}},
		\bibinfo {author} {\bibfnamefont {C.}~\bibnamefont {Persson}}, \ and\
		\bibinfo {author} {\bibfnamefont {Z.~J.}\ \bibnamefont {Wang}},\ }\href
	{\doibase 10.1016/j.cpc.2020.107760} {\bibfield  {journal} {\bibinfo
			{journal} {Comp. Phys. Commun.}\ }\textbf {\bibinfo {volume} {261}},\
		\bibinfo {pages} {107760} (\bibinfo {year} {2021})}\BibitemShut {NoStop}%
	\bibitem [{\citenamefont {Pizzi}\ \emph {et~al.}(2020)\citenamefont {Pizzi},
		\citenamefont {Vitale}, \citenamefont {Arita}, \citenamefont {Bluegel},
		\citenamefont {Freimuth}, \citenamefont {Geranton}, \citenamefont
		{Gibertini}, \citenamefont {Gresch}, \citenamefont {Johnson}, \citenamefont
		{Koretsune}, \citenamefont {Ibanez-Azpiroz}, \citenamefont {Lee},
		\citenamefont {Lihm}, \citenamefont {Marchand}, \citenamefont {Marrazzo},
		\citenamefont {Mokrousov}, \citenamefont {Mustafa}, \citenamefont {Nohara},
		\citenamefont {Nomura}, \citenamefont {Paulatto}, \citenamefont {Ponce},
		\citenamefont {Ponweiser}, \citenamefont {Qiao}, \citenamefont {Thoele},
		\citenamefont {Tsirkin}, \citenamefont {Wierzbowska}, \citenamefont
		{Marzari}, \citenamefont {Vanderbilt}, \citenamefont {Souza}, \citenamefont
		{Mostofi},\ and\ \citenamefont {Yates}}]{wannier90}%
	\BibitemOpen
	\bibfield  {author} {\bibinfo {author} {\bibfnamefont {G.}~\bibnamefont
			{Pizzi}}, \bibinfo {author} {\bibfnamefont {V.}~\bibnamefont {Vitale}},
		\bibinfo {author} {\bibfnamefont {R.}~\bibnamefont {Arita}}, \bibinfo
		{author} {\bibfnamefont {S.}~\bibnamefont {Bluegel}}, \bibinfo {author}
		{\bibfnamefont {F.}~\bibnamefont {Freimuth}}, \bibinfo {author}
		{\bibfnamefont {G.}~\bibnamefont {Geranton}}, \bibinfo {author}
		{\bibfnamefont {M.}~\bibnamefont {Gibertini}}, \bibinfo {author}
		{\bibfnamefont {D.}~\bibnamefont {Gresch}}, \bibinfo {author} {\bibfnamefont
			{C.}~\bibnamefont {Johnson}}, \bibinfo {author} {\bibfnamefont
			{T.}~\bibnamefont {Koretsune}}, \bibinfo {author} {\bibfnamefont
			{J.}~\bibnamefont {Ibanez-Azpiroz}}, \bibinfo {author} {\bibfnamefont
			{H.}~\bibnamefont {Lee}}, \bibinfo {author} {\bibfnamefont {J.-M.}\
			\bibnamefont {Lihm}}, \bibinfo {author} {\bibfnamefont {D.}~\bibnamefont
			{Marchand}}, \bibinfo {author} {\bibfnamefont {A.}~\bibnamefont {Marrazzo}},
		\bibinfo {author} {\bibfnamefont {Y.}~\bibnamefont {Mokrousov}}, \bibinfo
		{author} {\bibfnamefont {J.~I.}\ \bibnamefont {Mustafa}}, \bibinfo {author}
		{\bibfnamefont {Y.}~\bibnamefont {Nohara}}, \bibinfo {author} {\bibfnamefont
			{Y.}~\bibnamefont {Nomura}}, \bibinfo {author} {\bibfnamefont
			{L.}~\bibnamefont {Paulatto}}, \bibinfo {author} {\bibfnamefont
			{S.}~\bibnamefont {Ponce}}, \bibinfo {author} {\bibfnamefont
			{T.}~\bibnamefont {Ponweiser}}, \bibinfo {author} {\bibfnamefont
			{J.}~\bibnamefont {Qiao}}, \bibinfo {author} {\bibfnamefont {F.}~\bibnamefont
			{Thoele}}, \bibinfo {author} {\bibfnamefont {S.~S.}\ \bibnamefont {Tsirkin}},
		\bibinfo {author} {\bibfnamefont {M.}~\bibnamefont {Wierzbowska}}, \bibinfo
		{author} {\bibfnamefont {N.}~\bibnamefont {Marzari}}, \bibinfo {author}
		{\bibfnamefont {D.}~\bibnamefont {Vanderbilt}}, \bibinfo {author}
		{\bibfnamefont {I.}~\bibnamefont {Souza}}, \bibinfo {author} {\bibfnamefont
			{A.~A.}\ \bibnamefont {Mostofi}}, \ and\ \bibinfo {author} {\bibfnamefont
			{J.~R.}\ \bibnamefont {Yates}},\ }\href {\doibase 10.1088/1361-648X/ab51ff}
	{\bibfield  {journal} {\bibinfo  {journal} {J. Phys.-Condens. Matter}\
		}\textbf {\bibinfo {volume} {32}},\ \bibinfo {pages} {165902} (\bibinfo
		{year} {2020})}\BibitemShut {NoStop}%
	\bibitem [{\citenamefont {Sancho}\ \emph {et~al.}(1984)\citenamefont {Sancho},
		\citenamefont {Sancho},\ and\ \citenamefont {Rubio}}]{Green1984}%
	\BibitemOpen
	\bibfield  {author} {\bibinfo {author} {\bibfnamefont {M.~P.~L.}\
			\bibnamefont {Sancho}}, \bibinfo {author} {\bibfnamefont {J.~M.~L.}\
			\bibnamefont {Sancho}}, \ and\ \bibinfo {author} {\bibfnamefont
			{J.}~\bibnamefont {Rubio}},\ }\href {\doibase 10.1088/0305-4608/14/5/016}
	{\bibfield  {journal} {\bibinfo  {journal} {J. Phys. F-Met. Phys.}\ }\textbf
		{\bibinfo {volume} {14}},\ \bibinfo {pages} {1205} (\bibinfo {year}
		{1984})}\BibitemShut {NoStop}%
	\bibitem [{\citenamefont {Sancho}\ \emph {et~al.}(1985)\citenamefont {Sancho},
		\citenamefont {Sancho},\ and\ \citenamefont {Rubio}}]{Green1985}%
	\BibitemOpen
	\bibfield  {author} {\bibinfo {author} {\bibfnamefont {M.~P.~L.}\
			\bibnamefont {Sancho}}, \bibinfo {author} {\bibfnamefont {J.~M.~L.}\
			\bibnamefont {Sancho}}, \ and\ \bibinfo {author} {\bibfnamefont
			{J.}~\bibnamefont {Rubio}},\ }\href {\doibase 10.1088/0305-4608/15/4/009}
	{\bibfield  {journal} {\bibinfo  {journal} {J. Phys. F-Met. Phys.}\ }\textbf
		{\bibinfo {volume} {15}},\ \bibinfo {pages} {851} (\bibinfo {year}
		{1985})}\BibitemShut {NoStop}%
	\bibitem [{\citenamefont {Wu}\ \emph {et~al.}(2018)\citenamefont {Wu},
		\citenamefont {Zhang}, \citenamefont {Song}, \citenamefont {Troyer},\ and\
		\citenamefont {Soluyanov}}]{wanniertools}%
	\BibitemOpen
	\bibfield  {author} {\bibinfo {author} {\bibfnamefont {Q.}~\bibnamefont
			{Wu}}, \bibinfo {author} {\bibfnamefont {S.}~\bibnamefont {Zhang}}, \bibinfo
		{author} {\bibfnamefont {H.-F.}\ \bibnamefont {Song}}, \bibinfo {author}
		{\bibfnamefont {M.}~\bibnamefont {Troyer}}, \ and\ \bibinfo {author}
		{\bibfnamefont {A.~A.}\ \bibnamefont {Soluyanov}},\ }\href {\doibase
		10.1016/j.cpc.2017.09.033} {\bibfield  {journal} {\bibinfo  {journal} {Comp.
				Phys. Commun.}\ }\textbf {\bibinfo {volume} {224}},\ \bibinfo {pages} {405}
		(\bibinfo {year} {2018})}\BibitemShut {NoStop}%
	\bibitem [{\citenamefont {Luo}\ \emph {et~al.}(2020)\citenamefont {Luo},
		\citenamefont {Ji}, \citenamefont {Lu}, \citenamefont {Zhang},\ and\
		\citenamefont {Xiang}}]{FeB2-2020}%
	\BibitemOpen
	\bibfield  {author} {\bibinfo {author} {\bibfnamefont {W.}~\bibnamefont
			{Luo}}, \bibinfo {author} {\bibfnamefont {J.}~\bibnamefont {Ji}}, \bibinfo
		{author} {\bibfnamefont {J.}~\bibnamefont {Lu}}, \bibinfo {author}
		{\bibfnamefont {X.}~\bibnamefont {Zhang}}, \ and\ \bibinfo {author}
		{\bibfnamefont {H.}~\bibnamefont {Xiang}},\ }\href {\doibase
		10.1103/PhysRevB.101.195111} {\bibfield  {journal} {\bibinfo  {journal}
			{Phys. Rev. B}\ }\textbf {\bibinfo {volume} {101}},\ \bibinfo {pages}
		{195111} (\bibinfo {year} {2020})}\BibitemShut {NoStop}%
	\bibitem [{\citenamefont {Bistritzer}\ and\ \citenamefont
		{MacDonald}(2011)}]{TBG2011}%
	\BibitemOpen
	\bibfield  {author} {\bibinfo {author} {\bibfnamefont {R.}~\bibnamefont
			{Bistritzer}}\ and\ \bibinfo {author} {\bibfnamefont {A.~H.}\ \bibnamefont
			{MacDonald}},\ }\href {\doibase 10.1073/pnas.1108174108} {\bibfield
		{journal} {\bibinfo  {journal} {Proc. Natl. Acad. Sci. USA}\ }\textbf
		{\bibinfo {volume} {108}},\ \bibinfo {pages} {12233} (\bibinfo {year}
		{2011})}\BibitemShut {NoStop}%
	\bibitem [{\citenamefont {Song}\ \emph {et~al.}(2019)\citenamefont {Song},
		\citenamefont {Wang}, \citenamefont {Shi}, \citenamefont {Li}, \citenamefont
		{Fang},\ and\ \citenamefont {Bernevig}}]{PhysRevLett.123.036401}%
	\BibitemOpen
	\bibfield  {author} {\bibinfo {author} {\bibfnamefont {Z.}~\bibnamefont
			{Song}}, \bibinfo {author} {\bibfnamefont {Z.}~\bibnamefont {Wang}}, \bibinfo
		{author} {\bibfnamefont {W.}~\bibnamefont {Shi}}, \bibinfo {author}
		{\bibfnamefont {G.}~\bibnamefont {Li}}, \bibinfo {author} {\bibfnamefont
			{C.}~\bibnamefont {Fang}}, \ and\ \bibinfo {author} {\bibfnamefont {B.~A.}\
			\bibnamefont {Bernevig}},\ }\href {\doibase 10.1103/PhysRevLett.123.036401}
	{\bibfield  {journal} {\bibinfo  {journal} {Phys. Rev. Lett.}\ }\textbf
		{\bibinfo {volume} {123}},\ \bibinfo {pages} {036401} (\bibinfo {year}
		{2019})}\BibitemShut {NoStop}%
\end{thebibliography}
%

\clearpage

\begin{widetext}
\beginsupplement{}
\setcounter{section}{0}
\renewcommand{\thesubsection}{\Alph{subsection}}
\renewcommand{\thesubsubsection}{\alph{subsubsection}}

\section*{Appendix}

\subsection{Orbital-resolved band structures}
\label{sup:A}
The orbital-resolved band structures are presented in Fig.~\ref{fig:fatband}. We found that the low-energy bands near $E_F$ (\ie $-3 eV<E-E_F<1 eV$) mainly come from Fe-$d$ orbitals.

\begin{figure}[htbp]
	\includegraphics[width=4.5in]{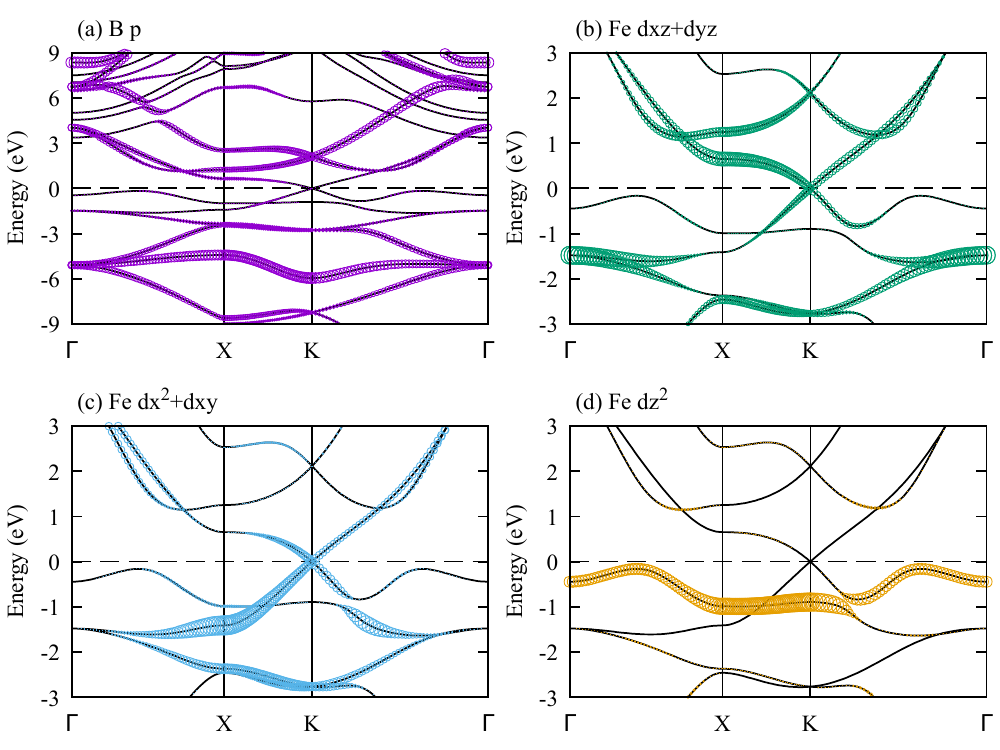}
	\caption{(color online).
		The orbital-resolved band structures of monolayer FeB$_2$ of (a)$p$ orbitals of B atoms, (b)$d_{xz}+d_{yz}$, (c)$d_{x^2-y^2}+d_{xy}$ and (d)$d_{z^2}$ orbitals of Fe atom.
	}\label{fig:fatband}
\end{figure}

\subsection{Low-energy effective Hamiltonians}
\label{sup:B}
Without SOC, the band crossing at $K$ belongs to two-dimensional irreducible representation (irrep) $G_{3}$ (labeled in the double point group of $C_{3v}$).
When considering SOC, it changes into $G_{4}{\oplus}G_{5}{\oplus}G_{6}$ irreps.
We choose the bases of $\{\ket{\uparrow} , \ket{\downarrow}\} \otimes \{\ket{d_{xz}+id_{yz}}, \ket{d_{xz}-id_{yz}}\}$ to construct the $4\times 4$ Hamiltonian.
The $\mathbf{k}\cdot\mathbf{p}$ Hamiltonian around $K$ is constrained by the double point group $C_{3v}$ and an anti-unitary symmetry $TC_{2z}$. Under these bases, the generators (\eg $C_{3z}$, $M_{y}$, and $C_{6z}T$) are represented as follows,

\begin{equation}\label{eq:C3z}
	\begin{split}
		C_{3z}&=
		\left(
		\begin{array}{cccc}   %
			-1 &  0  &   0   &   0     \\  %
			0 &  e^{i\pi/3}   &  0  &   0    \\  %
			0  &  0   &   e^{-i\pi/3}  &    0  \\  %
			0  &  0  &   0    &  -1  \\  %
		\end{array}
		\right),
	\end{split}
\end{equation}

\begin{equation}\label{eq:My}
	\begin{split}
		M_{y}&=
		\left(
		\begin{array}{cccc}   %
			0 &  0  &   0   &   1     \\  %
			0 &  0   &  1  &   0    \\  %
			0  &  -1   &   0  &    0  \\  %
			-1  &  0  &   0    &  0  \\  %
		\end{array}
		\right),
	\end{split}
\end{equation}

\begin{equation}\label{eq:C6zT}
	\begin{split}
		C_{6z}T&=
		\left(
		\begin{array}{cccc}   %
			0 &  0  &   0   &   i     \\  %
			0 &  0   &  -e^{i\pi/6}  &   0    \\  %
			0  &  e^{-i\pi/6}   &   0  &    0  \\  %
			i  &  0  &   0    &  0  \\  %
		\end{array}
		\right){\cal K},
	\end{split}
\end{equation}
where ${\cal K}$ denotes the complex conjugation.
Using the theory of invariants, we constructed the $\mathbf{k}\cdot\mathbf{p}$ effective Hamiltonian below, 
\begin{equation}\label{eq:kp-sm}
	\begin{split}
		H_{K}^{so}(\vec{k})&=
		\left(
		\begin{array}{cccc}   %
			M_{1}(\vec{k})+M_{2}(\vec{k}) &    &      &   \dagger     \\  %
			Ak_{-} &  M_{1}(\vec{k})-M_{2}(\vec{k})   &    &       \\  %
			iBk_{+}  &  iC(\vec{k})   &   M_{1}(\vec{k})-M_{2}(\vec{k})  &      \\  %
			iM_{3}(\vec{k})  &  iBk_{+}  &   Ak_{-}    &  M_{1}(\vec{k})+M_{2}(\vec{k})  \\  %
		\end{array}
		\right),
	\end{split}
\end{equation}
where $k_{\pm}=k_{x}\pm ik_{y}$, $C(\vec{k})=C_{1}k_{-}+C_{2}k_{+}^{2}$, and $M_{\alpha=1,2,3}(\vec{k})=E_{\alpha}+F_{\alpha}k_{\bot}^{2}$ with $k_{\bot}^{2}=k_{x}^{2}+k_{y}^{2}$.
On the other hand, the simplest two-band $\mathbf{k}\cdot\mathbf{p}$ Hamiltonian under the two bases of $G_{6}$ irrep is also derived as,
\begin{equation}\label{eq:kp-g6-sm}
	\begin{split}
		H'_{K}(\vec{k})&=
		\left(
		\begin{array}{cc}   %
			M_{1}(\vec{k})-M_{2}(\vec{k})   &  -iC_{1}k_{+}-iC'_{2}k_{-}^{2} \\  %
			iC_{1}k_{-}+iC'_{2}k_{+}^{2}   &   M_{1}(\vec{k})-M_{2}(\vec{k})  \\  %
		\end{array}
		\right).
	\end{split}
\end{equation}

Then, we will prove there are only four Dirac points in the above two-band Hamiltonian. With the condition $k_x=ak_y$ ($a\neq0$), we will get solutions ($k_x=(a^2(3-a^2)C_1/C'_2\pm\sqrt{-(3a^3-a)^2(C_1/C'_2)^2})/(1+a^2)^2,k_y=k_x/a$). To get real roots of the solutions, $\sqrt{-(3a^3-a)^2(C_1/C'_2)^2}$ should be zero. If $C_1/C'_2\neq0$ and $a\neq0$, then $3a^2-1=0$, thus $a=\pm\sqrt{3}/3$. Taking $k_x=\pm{k_y/\sqrt{3}}$ into the equation, we can get the positions of two Dirac points, which are ($C_1/2C'_2, \pm\sqrt{3}C_1/2C'_2$). Combining the results of conditions $k_x=0$, $k_y=0$ and $k_x=ak_y$ ($a\neq0$), there are four Dirac points in two-band $\mathbf{k}\cdot\mathbf{p}$ Hamiltonian of $G_{6}$ irrep. One Dirac point is located at ($0, 0$), and three $C_3$-related Dirac points are located at ($-C_1/C'_2,0$) and ($C_1/2C'_2, \pm\sqrt{3}C_1/2C'_2$). These three Dirac points are connected by $C_3$ symmetry, and their distances from $K$ are $d_0=|C_1/C'_2|$. Thus, with the $|C_1/C'_2|$ decreases, the Dirac points will come close to the $K$ point. If $C_1$ decrease to zero, the eigenvalues of Eq.~(\ref{eq:kp-g6-sm}) are $E_{\pm}=M_{1}(\vec{k})-M_{2}(\vec{k})\pm|C'_{2}|(k_x^2+k_y^2)=E_{1}-E_{2}+(F_{1}-F_{2}\pm|C'_2|)k_{\bot}^{2}$, indicating a quadratic dispersion around $K$, corresponding to a double Dirac point with $2\pi$ Berry phase.

\begin{figure}[htbp]
	\includegraphics[width=4.5in]{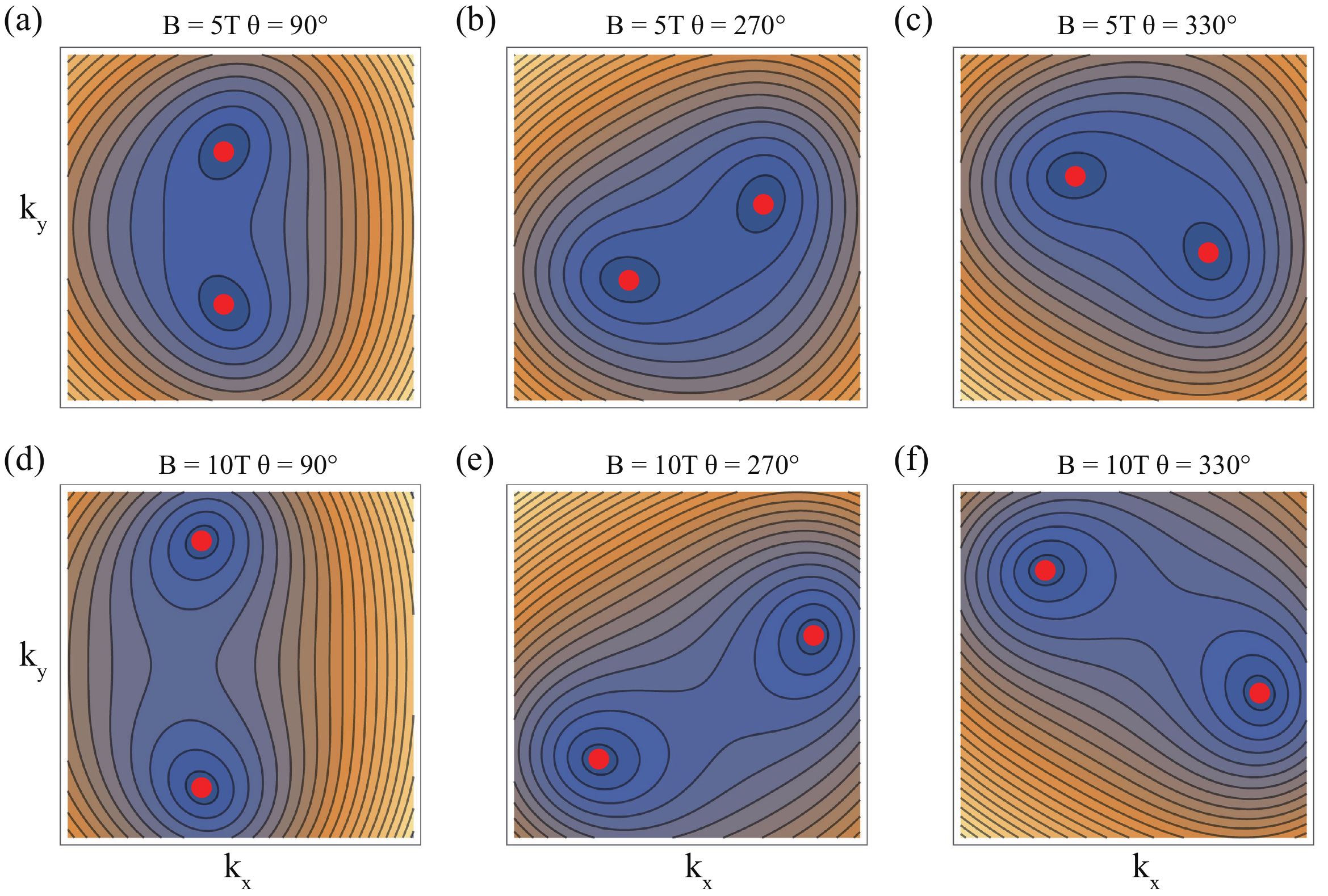}
	\caption{(color online).The iso-energy-gap contours in rectangle area with $5\times10^{-4}$ \AA$^{-1}$ around $K$ under (a-c)5T and (d-f)10T with different directions of the in-plane magnetic field. The positions of Dirac nodes are marked by red dots. The angle between the magnetic field and the $x$ axis located in the $xy$ plane is marked by $\theta$.}
	\label{fig:mag_in_plane}
\end{figure}

\subsection{Zeeman splitting under magnetic field}
\label{sup:C}
The Zeeman's coupling is obtained as,
\begin{equation}\label{eq:kp-zeeman}
	\begin{split}
		H^{Z}(\vec{B})&=
		\mu_B\left(
		\begin{array}{cccc}   %
			g_{\parallel}^{\frac{3}{2}}B_z &          & &\dagger      \\  %
			ig'_{\perp}B_{-}  &  -g_{\parallel}^{\frac{1}{2}}B_z   & &\\  %
			g''_{\perp}B_{+}   &   g_{\perp}^{\frac{1}{2}}B_{-}   &  g_{\parallel}^{\frac{1}{2}}B_z  &  \\  %
			0  &    g''_{\perp}B_{+}    & ig'_{\perp}B_{-} &  -g_{\parallel}^{\frac{3}{2}}B_z  \\  %
		\end{array}
		\right),
	\end{split}
\end{equation}
where $B_{\pm}=B_{x}\pm iB_{y}$ and $\mu_B=\frac{e\hbar}{2m_e}$ is Bohr magneton. 
Under in-plane magnetic field (which keeps $TC_{2z}$), the double Dirac point splits into two Dirac points with identical Berry phase $\pi$, as shown in Fig.~\ref{fig:mag}(a,b) and Fig.~\ref{fig:mag_in_plane}.
Besides, we consider in-plane magnetic fields with different strengths and directions, and find that the Dirac nodes move away from $K$ when increasing the magnetic field.
Then we consider the external magnetic field in the $z$ direction (breaking $TC_{2z}$).
This Zeeman field will open an energy gap at both K points, giving rise to a Chern insulator with $C=2$.

\subsection{Three-orbital tight-binding model}
\label{sup:D}

\begin{figure}[htbp]
	\includegraphics[scale=0.3]{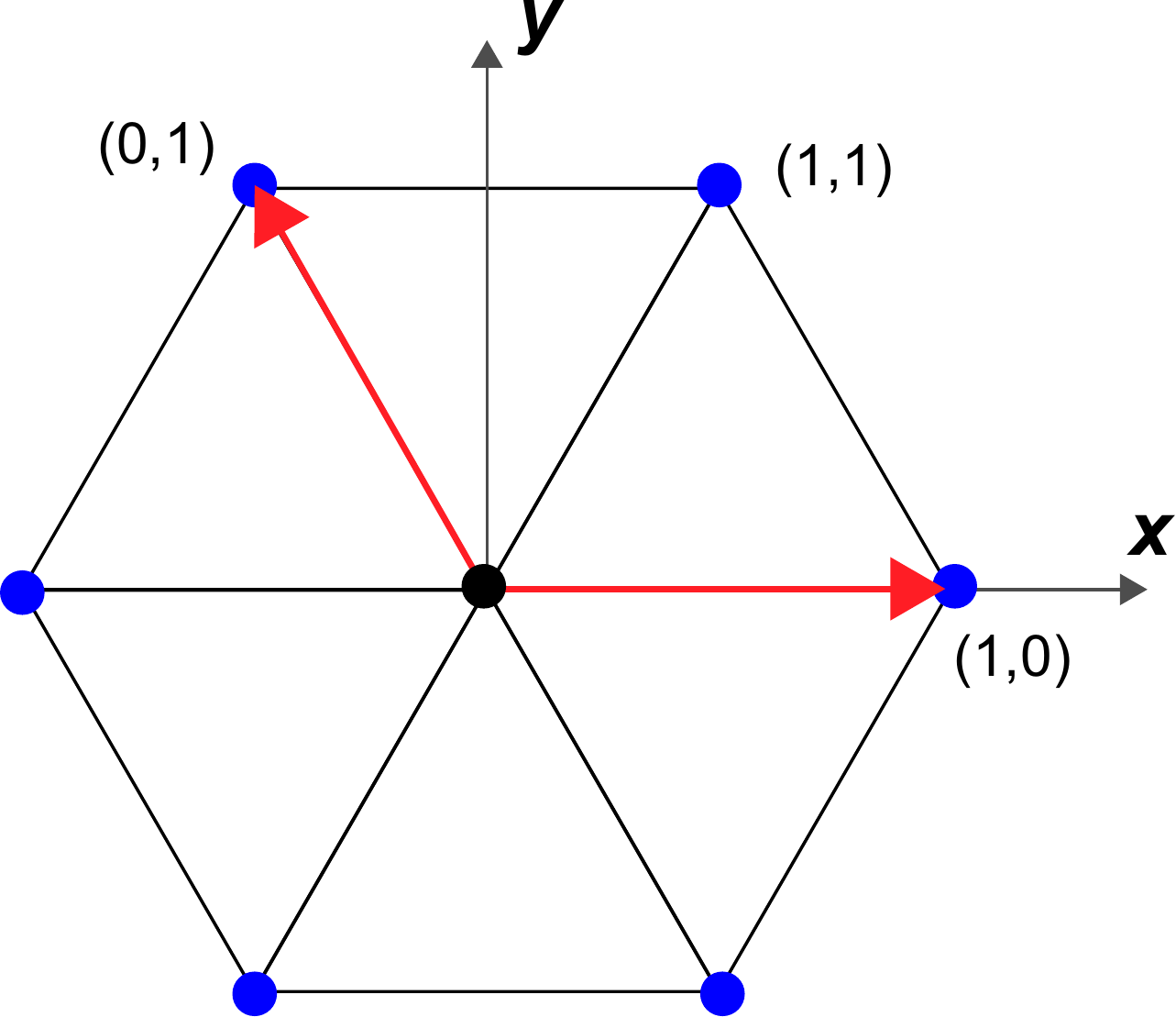}
	\caption{The lattice of the tight-binding model with a lattice constant $a$. The 1a Wyckoff site is (0,0) with respect to the lattice vectors, denoted by two red-colored arrows.}
	\label{fig:tri}
\end{figure}

To capture the Dirac points at K points, the elementary band representations have to be $\{A_1@1a$ and $E_1@1a\}$ in the absence of SOC.
Thus we choose $d_{xz}, d_{yz}, d_{z^2}$ orbitals of Fe atom to build the tight-binding model. The state $|d_{\alpha}, (00) \rangle$ defines the orbital $d_{\alpha}$ in the unit cell (mn) with respect to lattice rectors (lattice constant $a$), as shown in Fig.~\ref{fig:tri}.
Only the nearest neighbor hoppings are considered.

Consider the spin degree of freedom, the $6\times 6$ tight-binding (tb) Hamiltonian is given below,
\begin{equation}
	H_{tb}=\sigma_0\otimes H(\vec k)+ H_{so}(\vec k),~H_{so}(\vec k)=\left(\begin{array}{cc}   
		\Lambda_0(\vec k)& \Lambda_1(\vec k) \\
		-\Lambda^T_1(-\vec k) &\Lambda^T_0(-\vec k) \\
	\end{array}
	\right)
\end{equation}
The $ H(\vec k), \Lambda_0(\vec k), \Lambda_1(\vec k)$ are $3\times3$ matrices with $\vec k$ given in units of $1/a$.  
The up-triangle elements of the Hermitian matrix $H(\vec k)$ without SOC are given as follows,
\begin{equation}
	\begin{split}
		H_{11}(\vec k) &= \epsilon_1 + 2t_1\cos{k_x} + ( t_1+3t_2)\cos{\frac{k_x}{2}}\cos{\frac{\sqrt{3}}{2}k_y} \\
		H_{22}(\vec k) &= \epsilon_1 + 2t_2\cos{k_x} + (3t_1+ t_2)\cos{\frac{k_x}{2}}\cos{\frac{\sqrt{3}}{2}k_y} \\
		H_{33}(\vec k) &= \epsilon_2 + 2t_3[\cos{k_x}+\cos{(\frac{k_x}{2}+\frac{\sqrt{3}}{2}k_y)}+\cos{(-\frac{k_x}{2}+\frac{\sqrt{3}}{2}k_y)}] \\
		H_{12}(\vec k) &= \sqrt{3}(t_2-t_1)\sin{\frac{k_x}{2}}\sin{\frac{\sqrt{3}}{2}k_y} \\
		H_{13}(\vec k) &= it_4[ 2\sin{k_x} + \sin{(\frac{k_x}{2}+\frac{\sqrt{3}}{2}k_y)} - \sin{(-\frac{k_x}{2}+\frac{\sqrt{3}}{2}k_y)} ] \\
		H_{23}(\vec k) &= i\sqrt{3}t_4[ \sin{(\frac{k_x}{2}+\frac{\sqrt{3}}{2}k_y)} + \sin{(-\frac{k_x}{2}+\frac{\sqrt{3}}{2}k_y)} ] \\
		{\rm with~}&\epsilon_1 = \langle d_{xz},(00) | H | d_{xz}, (00) \rangle ,
		\epsilon_2 = \langle d_{z^2},(00) | H | d_{z^2}, (00) \rangle,
		t_1 = \langle d_{xz},(00) | H | d_{xz}, (10) \rangle, \\
		&t_2 = \langle d_{yz},(00) | H | d_{yz}, (10) \rangle ,
		t_3 = \langle d_{z^2},(00) | H | d_{z^2}, (10) \rangle ,
		t_4 = \langle d_{xz},(00) | H | d_{z^2}, (10) \rangle .
	\end{split}
\end{equation}

The spin-orbit coupling terms of $\Lambda_0(\vec k)$ and $\Lambda_1(\vec k)$ are derived as,
\begin{equation}
	\begin{split}
		\Lambda_0(\vec k)& = \begin{pmatrix}
			0 & i\lambda_0 & 0 \\
			-i\lambda_0 & 0 & 0 \\
			0 & 0 & 0
		\end{pmatrix},~
		\Lambda_1(\vec k) = \begin{pmatrix}
			a(\vec k) & c(\vec k)  & 0 \\
			c(\vec k) & b(\vec k) & 0 \\
			0 & 0 & 0
		\end{pmatrix},  \\
		a (\vec k)& = 2i\lambda_1\sin{k_x} + 2ie^{-i\frac{\pi}{3}}(\frac{\lambda_1}{4}+\frac{3\lambda_2}{4})[\sin{(\frac{k_x}{2}+\frac{\sqrt{3}k_y}{2})}+e^{-i\frac{\pi}{3}}\sin{(-\frac{k_x}{2}+\frac{\sqrt{3}k_y}{2})}] \\
		b (\vec k)& = 2i\lambda_2\sin{k_x} + 2ie^{-i\frac{\pi}{3}}(\frac{3\lambda_1}{4}+\frac{\lambda_2}{4})[\sin{(\frac{k_x}{2}+\frac{\sqrt{3}k_y}{2})}+e^{-i\frac{\pi}{3}}\sin{(-\frac{k_x}{2}+\frac{\sqrt{3}k_y}{2})}] \\
		c (\vec k)& = i\frac{\sqrt{3}}{2}(\lambda_1-\lambda_2)[ e^{-i\frac{\pi}{3}}\sin{(\frac{k_x}{2}+\frac{\sqrt{3}k_y}{2})} + e^{i\frac{\pi}{3}}\sin{(-\frac{k_x}{2}+\frac{\sqrt{3}k_y}{2})} ] \\
		{\rm with~}&i\lambda_0= \langle d_{xz\uparrow},(00) | H | d_{yz\uparrow}, (00) \rangle, \\
		&\lambda_1 = \langle d_{xz\uparrow},(00) | H | d_{xz\downarrow}, (10) \rangle ,
		\lambda_2 = \langle d_{yz\uparrow},(00) | H | d_{yz\downarrow}, (10) \rangle.\\
	\end{split}
\end{equation}

\begin{figure}[htbp]
	\includegraphics[width=4.5in]{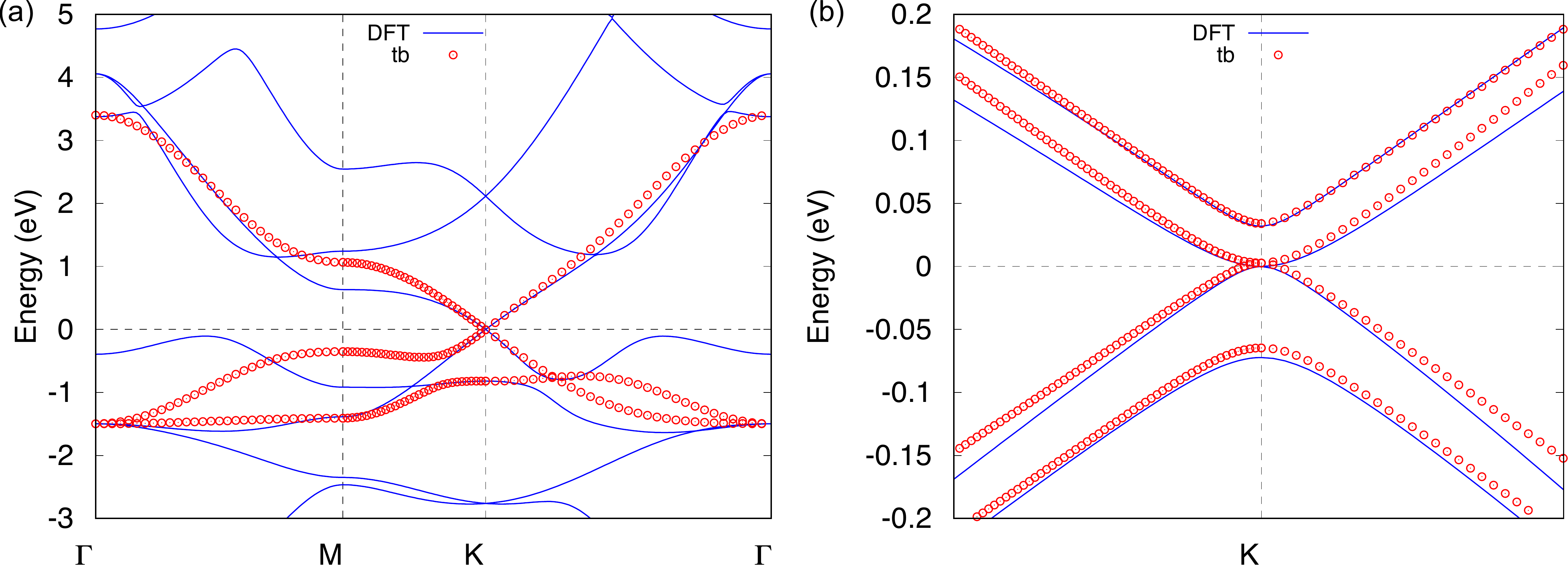}
	\caption{The comparison between the DFT bands and the energy bands of the tight-binding model without (a) and with (b) SOC. The parameters are given in Table~\ref{table:tbparam}. }
	\label{fig:tbfit}
\end{figure}

\begin{table}[htbp]
	\setlength{\tabcolsep}{1.0mm}
	\setlength{\extrarowheight}{2pt}
	\caption{The parameters in the tight-binding Hamiltonian are real and given in units of eV.}
	\begin{tabular}{ccccccccc}
		\hline
		\hline
		$\epsilon_1$ & 	$\epsilon_2$ &$t_1$  & $t_2$ & $t_3$ & $t_4$ & $\lambda_1$&$\lambda_2$ &$\lambda_0$ \\
		\hline 
		-0.5036 & 0.5864 & -0.475 & 0.1434 & 0.4689 & -0.3500 & -0.0290 & -0.0100 & 0.008\\
		\hline
		\hline
	\end{tabular}
	\label{table:tbparam}
\end{table}

By fitting the energy bands of the DFT calculation without SOC, the parameters of $\epsilon_1, \epsilon_2, t_1,t_2,t_3$ and $t_4$ are obtained. The three parameters $\lambda_{0,1,2}$ are obtained by fitting the DFT bands with SOC. These parameters are listed in Table~\ref{table:tbparam}, and the corresponding band structures are shown in Fig.~\ref{fig:tbfit}.

\subsection{The band structures of LDA+U method}
\label{sup:E}
We calculate the LDA+U and LDA+U+SOC band structures, and find that the quadratic band crossing points are preserved in LDA+U+SOC results, as shown in Fig.\ref{fig:band_plusu}. We would like to emphasize here that the former LDA+U technic we used is based on Dudarev\ea's work\cite{ldau}. Only the density-related terms are counted in the approximation. However, the crucial inter-orbital correlation terms are not considered in these calculations. Thus, we think the normal LDA+U method is not a suitable method to explore the physics here.
\begin{figure}[htbp]
	\includegraphics[width=6.5in]{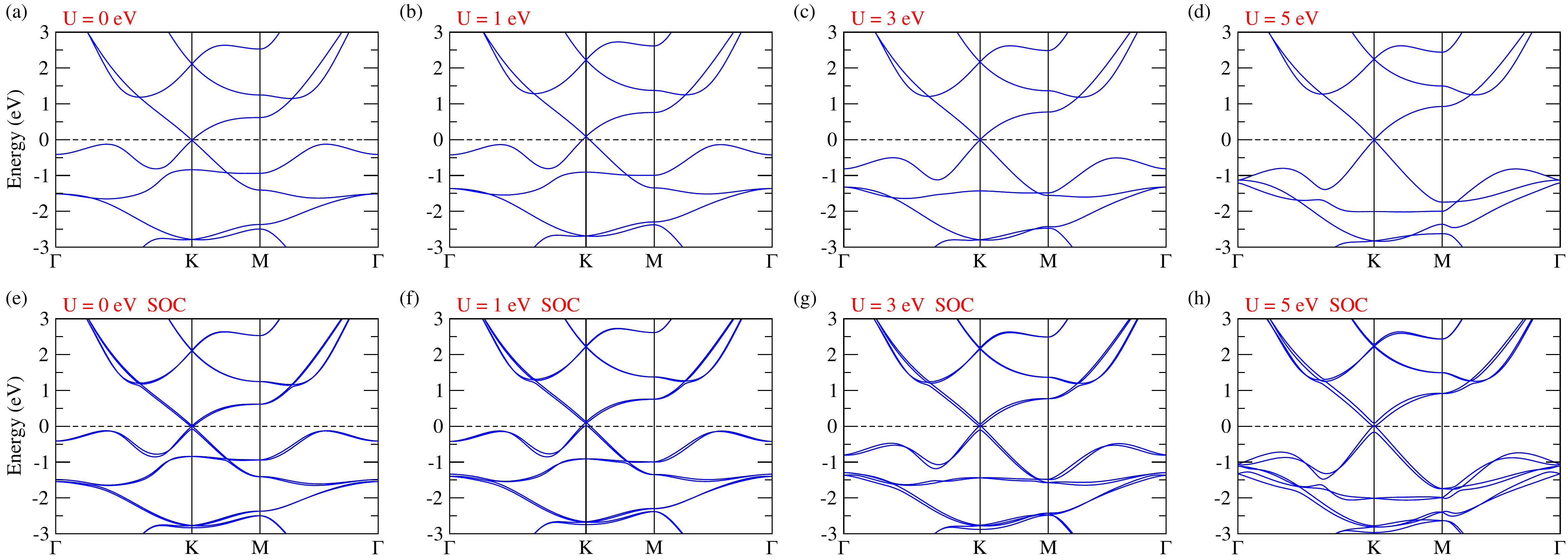}
	\caption{(a-d)The band structures of FeB$_2$ monolayer with Hubbard U of 0, 1, 3, and 5 eV. (e-h)The LDA+U+SOC band structures of FeB$_2$ monolayer. The quadratic band crossing points are located at $K$.}
	\label{fig:band_plusu}
\end{figure}

\clearpage
\end{widetext}
\end{document}